\documentclass[twocolumn,times]{aastex631}

\usepackage{graphicx}
\usepackage[flushleft]{threeparttable}
\usepackage{blindtext}
\usepackage{amsmath}
\usepackage{mathtools}
\usepackage{multirow}
\usepackage{comment}
\turnoffeditone

\begin{document}
\shortauthors{Jung et al.}
\def\nar{New Astron.}
\def\na{New Astron.}

\title{\Large Local Analogs of Potential Ionizers of the Intergalactic Medium: \\Compact Star-Forming Galaxies with Intense \ion{C}{4} $\lambda$1550 Emission}

\correspondingauthor{Intae Jung}
\email{ijung@stsci.edu}

\author[0000-0003-1187-4240]{Intae Jung}
\affiliation{Space Telescope Science Institute, 3700 San Martin Drive Baltimore, MD 21218, United States}

\author[0000-0002-5269-6527]{Swara Ravindranath}
\affiliation{Astrophysics Science Division, NASA Goddard Space Flight Center, 8800 Greenbelt Road, Greenbelt, MD 20771, USA}
\affiliation{Center for Research and Exploration in Space Science and Technology II, Department of Physics, Catholic University of America, 620 Michigan Ave N.E., Washington DC 20064, USA}

\author[0000-0002-6790-5125]{Anne E. Jaskot}
\affiliation{Department of Physics and Astronomy, Williams College, Williamstown, MA, 01267, USA}

\author[0000-0001-7113-2738]{Henry C. Ferguson}
\affiliation{Space Telescope Science Institute, 3700 San Martin Drive Baltimore, MD 21218, United States}

\author[0000-0003-4372-2006]{Bethan L. James}
\affiliation{AURA for ESA, Space Telescope Science Institute, 3700 San Martin Drive, Baltimore, MD 21218, USA}

\begin{abstract}
We performed spectroscopic analyses of five local compact star-forming galaxies (CSFGs) with extremely high [\ion{O}{3}]/[\ion{O}{2}] (O$_{32}$) ratios ($>$\,$20$). These targets remarkably share similar properties with high-redshift \ion{C}{4} emitters at $z>6$: high H$\beta$ equivalent widths (EWs $>200$\AA), extreme O$_{32}$ ratios, low metallicities (12+log(O/H) $\lesssim7.8$), low C/O abundances (log(C/O) $<-0.6$), and high ionization conditions (log$U>-2$). The UV spectra were acquired using the Hubble Space Telescope's (HST) Cosmic Origins Spectrograph (COS) and Space Telescope Imaging Spectrograph (STIS). We have identified a wealth of rest-frame UV emission lines (\ion{C}{4}, \ion{He}{2}, \ion{O}{3}], \ion{C}{3}]) in the HST spectra. Notably, all our targets show intense \ion{C}{4} emission lines with rest-frame EWs $>10$\AA, indicative of hard ionizing radiation. The rest-frame UV emission line diagnostics disfavor an AGN and could be consistent with significant shock contributions to the source of ionizing radiation. Four of our targets show high \ion{C}{4}/\ion{C}{3}] ratios ($\geq$\,$1.4$), suggestive of strong Lyman-continuum leakage (LyC escape fraction, $f_{\rm esc,LyC}>10$\%) from these sources. This is consistent with their Ly$\alpha$-inferred LyC escape fractions ($f_{\rm esc,LyC}=9$\,--\,$31$\%).  We derive relative C/O abundances from our sources, showing log(C/O) values from $-1.12$ to $-0.61$, comparable to those of reionization-era galaxies at $z\gtrsim6$. The properties of the CSFGs, particularly their intense \ion{C}{4} emission and high O$_{32}$ ratios, which suggest significant LyC escape fractions, are similar to those of the reionization-era \ion{C}{4} emitters. These similarities reinforce the hypothesis that these CSFGs are the closest analogs of significant contributors to the reionization of the intergalactic medium.
\end{abstract}
\submitjournal{the Astrophysical Journal}

\section{Introduction}
The epoch of reionization (EoR) marks a pivotal period in the history of the universe when the first galaxies formed and ionized the intergalactic medium (IGM). Understanding the sources and mechanisms of Lyman-continuum (LyC) photon escape from galaxies is crucial for elucidating the reionization process, although the contribution from active galactic nuclei (AGNs) may not be negligible \citep[e.g.,][]{Robertson2022a}. In particular, low-mass, low-metallicity star-forming galaxies during the EoR have been recognized as significant contributors to this transformative cosmic event \citep[e.g.,][]{Wise2014a, Paardekooper2015a, Xu2016a, Anderson2017a, Atek2024a, Simmonds2024a}. Crucially, the relative contributions to the total ionizing photon budget between bright and faint sources must be imprinted in the temporal and spatial evolution of reionization \citep[e.g.,][]{Robertson2015a, Finkelstein2019b, Naidu2020a, Kannan2022a}.

High-ionization ultraviolet (UV) emission lines serve as key diagnostics for the properties of the interstellar medium (ISM) and the potential for these galaxies to contribute to reionization \citep[e.g.,][]{Stark2015a, Stark2015b, Berg2016a, Berg2019a, Mainali2017a, Senchyna2017a, Mingozzi2022a}. However, observing galaxies in the reionization era has been challenging due to the shift of important features into the infrared and the long exposure times on large ground-based telescopes needed to obtain rest-frame UV spectra \citep{Mainali2017a, Stark2017a, Hutchison2019a, Topping2021a}. JWST made a great leap on this purpose, providing its unique access to such redshifted spectra; however, direct measurements of LyC radiation from reionization-era galaxies are inherently limited due to IGM attenuation.  Instead, local analogs of high-redshift galaxies serve as excellent targets for detailed examination of the ionizing radiation escape mechanisms from galaxy populations that were likely prevalent during reionization \citep[e.g.,][]{Izotov2016a, Izotov2016b, Izotov2018a, Izotov2018b, Berg2019a, Berg2022a, Wang2019a, Izotov2021a, James2022a, Flury2022a, Flury2022b, Mingozzi2024a}. A distinctive feature has emerged, for example, from the study of intense \ion{C}{4} emission in local analogs to these distant galaxies, particularly those with low mass ($M_{*}\sim10^6{\rm M}_{\odot}$), low metallicity (12+log(O/H) $\lesssim$ 7.5), and intense star formation \citep[e.g.,][]{Berg2016a, Berg2019a, Berg2022a, Senchyna2017a, Senchyna2019a, Wofford2021a, Schaerer2022a, Izotov2024a}. Notably, the link between strong \ion{C}{4} emissions and LyC leakage, observed in LyC-leaking local analogs, underscores the potential of these galaxies as key to understanding reionization \citep{Schaerer2022a}.

Particularly, recent observations with the James Webb Space Telescope (JWST) have revealed remarkable \ion{C}{4} $\lambda$1550 emission from high-redshift galaxies, suggesting that intense radiation fields may be much more common in the reionization era \citep[e.g.,][]{Bunker2023a, Fujimoto2023a, Tang2023a, Topping2024a, Topping2024b, Witstok2024a, Castellano2024a, Kumari2024a, Curti2024a}. These findings highlight the potential of these galaxies to dominate the ionizing background during reionization. However, the exact origin of the high-ionization lines—whether from very low-metallicity stellar populations or other extreme sources (e.g., AGNs, Very Massive Stars, and shocks)—remains an area of active research \citep[e.g.,][]{Castellano2024a, Topping2024a, Topping2024b, Schaerer2024a, Mazzolari2024b}.

More detailed investigations of \ion{C}{4} emitters have been undertaken in local and intermediate-redshift galaxies, up to $z\sim3$, to understand the production of their intense ionizing fields.  Young, massive stars can produce high-energy photons capable of ionizing species like \ion{C}{4} ($>$47.9 eV) and \ion{He}{2} ($>$54.4 eV). Theoretical models and observations have shown that metal-poor, high-mass stars can contribute to the hard ionizing spectra needed to explain the high-ionization lines seen \citep[e.g.,][]{Shapley2003a, Leitherer2011a, Crowther2016a, Steidel2016a, Nakajima2018a, Smith2023a, Cameron2024a, Wofford2023a, Schaerer2024a}. In some cases, the presence of an AGN can also contribute to the high-ionization state \citep[e.g.,][]{Steidel2011a, Hainline2011a, Alexandroff2013a}. In addition to photoionization from energetic sources, shock ionization can contribute to the emergence of these high-ionization lines \citep[e.g.,][]{Allen2008a, Jaskot2016a, Alarie2019a}. Thus, distinguishing between ionization sources requires careful analysis of emission line ratios using photoionization models \citep{Feltre2016a, Gutkin2016a, Hirschmann2023a, Mazzolari2024a}.

While it is evident that \ion{C}{4} emitters produce high-energy ionizing photons, these photons must escape the dense ISM of galaxies to contribute to the reionization of the IGM. High LyC escape fractions, in general, are expected to be associated with low-metallicity environments, highly ionized ISM with young stellar populations, lack of dust, and compact morphologies, suggesting that certain physical conditions are conducive to LyC leakage \citep[e.g.,][]{Heckman2001a, Jaskot2013a, Nakajima2014a, Chisholm2018a, Jaskot2019a, Kakiichi2021a, Kimm2022a}. Specifically, factors that influence LyC escape include the geometry and porosity of the ISM, the presence of low-density channels created by stellar feedback, and the overall star formation rate, which has been theoretically predicted and supported by observational studies \citep[e.g.,][]{Kimm2014a, Cen2015a, Ma2015a, Paardekooper2015a, Jaskot2016a, Rivera-Thorsen2019a, Komarova2021a, Hu2023a, Kim2023a, Witten2024a, Choustikov2024a, Amorin2024a, Jung2024b}. Despite such complex nature of LyC escape, observational studies have attempted to infer LyC escape fractions in galaxies from indirect indicators, including, for example, the escape of Ly$\alpha$, the [\ion{O}{3}]/[\ion{O}{2}] line ratio, \ion{Mg}{2} emission, rest-UV $\beta$ slope, UV absorption lines, and galaxy sizes \citep[e.g.,][]{Dijkstra2016a, Jaskot2013a, Verhamme2015a, Henry2018a, Izotov2018a, Izotov2020a, Chisholm2018a, Gazagnes2018a, Chisholm2020a, Chisholm2022a, Saldana-Lopez2022a, Xu2022a, Xu2023a}. Particularly, the \ion{C}{4}/\ion{C}{3}] line ratio is linked to strong LyC leakers with an escape fraction of $>$10\% based on observations of intense \ion{C}{4} emitters \citep{Schaerer2022a, Saxena2022b, Kramarenko2024a}. Thus, emerging \ion{C}{4} emission line measurements from reionization-era galaxies make them key features to understanding the mechanisms that enable the escape of ionizing photons.

The motivation for this study stems from the need to explore more extreme cases of local analogs (e.g., intense UV emission lines and extreme [\ion{O}{3}]/[\ion{O}{2}] ratios) that mostly resemble dominant ionizers of the IGM during the reionization epoch, being witnessed from recent JWST observations \citep[e.g.,][]{Bunker2023a, Topping2024a, Castellano2024a, Kumari2024a, Curti2024a}. We present Hubble Space Telescope (HST) UV spectroscopic observations of five local compact star-forming galaxies (CSFGs) with extremely high [\ion{O}{3}]/[\ion{O}{2}] (O$_{32}$) ratios ($>20$) that show intense \ion{C}{4} emission. We investigate their ionizing sources and physical properties, including the ionization conditions, potential LyC escape, and chemical abundances, to better understand the processes that govern the EoR. Understanding these extreme local analogs can provide critical insights into the physical conditions that promote LyC escape and the overall contribution of such galaxies to reionization. 

This paper is organized as follows. Section 2 provides details of our targets and spectroscopic data. Section 3 describes the emission line measurements and the analysis of chemical abundances. In Section 4, we discuss ionization sources based on key emission line diagnostics, characterize physical conditions, diagnose the potential for escaping LyC photons, and interpret the chemical abundances of our sources. Finally, Section 5 summarizes the findings and discusses their implications for understanding reionization-era galaxies.  We assume the Planck cosmology \citep{Planck-Collaboration2016a} in this work: $H_0$ = 67.8\,km\,s$^{-1}$\,Mpc$^{-1}$, $\Omega_{\text{M}}$ = 0.308, and $\Omega_{\Lambda}$ = 0.692. The error measurements in this paper quote 1$\sigma$ uncertainties (or central 68\% confidence ranges) unless stated otherwise.

\begin{deluxetable*}{cccccccc}
\tabletypesize{\footnotesize}
\tablecaption{Targets and HST Observation Summary} 
\tablehead{\colhead{ID} & \colhead{R.A. (J2000.0)} & \colhead{Decl. (J2000.0)}  & \colhead{$z$} & \colhead{$M_{\rm UV}$} & \colhead{$R_{\rm eff}$} & \colhead{COS/FUV G140L Exposure} & \colhead{STIS/NUV-MAMA G230L Exposure}\\
\colhead{} & \colhead{(deg)} & \colhead{(deg)} & \colhead{} &  \colhead{(mag)} & \colhead{(pc)} & {(sec)} & {(sec)}\\
\colhead{(1)} & \colhead{(2)} & \colhead{(3)} & \colhead{(4)} &  \colhead{(5)} & \colhead{(6)} & {(7)} & {(8)}}
\startdata
{J0159+0751} & {01:59:52.75} & {+07:51:48.90} & {0.0611} & {$-16.3$} & {157} & {7310} & {4052} \\
{J1032+4919} & {10:32:56.72} & {+49:19:47.24} & {0.0442} & {$-16.1$} & {106} & {4794} & {1868}  \\
{J1205+4551} & {12:05:3.55} & {+45:51:50.94} & {0.0654}  & {$-16.6$} & {150} & {7519} & {4028}  \\
{J1355+4651} & {13:55:25.64} & {+46:51:51.34} & {0.0281} & {$-14.9$} & {83} & {7729} & {4508}  \\
{J1608+3528} & {16:08:10.36} & {+35:28:9.30} & {0.0327}  & {$-15.5$} & {105} & {4624} & {1760} \\
\enddata
\tablecomments{Columns: (1) Object ID, (2) Right ascension, (3) Declination, (4) Redshift from \cite{Izotov2017a}, (5) Absolute UV magnitude, measured in a wavelength window of rest-frame $1450$\,--\,$1550$\AA\ from COS/FUV G140L spectra, (6) Half-light radius, estimated from HST/COS NUV acquisition images using the PHOTUTILS package \citep{Bradley2023a}, (7) Exposure time in COS/FUV G140L observation, (8) Exposure time in STIS/NUV-MAMA G230L observation.}
\label{tab:targets}
\end{deluxetable*}

\section{Sample and Data}
\subsection{High [\ion{O}{3}]/[\ion{O}{2}] Galaxies}
Our sample consists of five CSFGs at low redshifts ($z = 0.028-0.065$). The targets were first introduced in \cite{Izotov2017a}.  These galaxies are selected for their extreme ionization conditions, indicated by their very high [\ion{O}{3}]$\lambda$5007/[\ion{O}{2}]$\lambda$3727 (O$_{32}$) ratios, which range from 23 to 49. Such high O$_{32}$ ratios imply exceptionally high ionization parameters (log $U > -2$). These CSFGs are characterized by low metallicities (12+log(O/H) $\leq7.8$). They exhibit high H$\beta$ rest-frame equivalent widths (EWs) between 220 and 510\AA, indicative of vigorous star formation and young stellar populations with ages less than 3 million years. Also, these galaxies have compact sizes (roughly $80$ to $160$ pc in the half-light radii) and relatively low stellar masses ($10^6$ to $10^7$ $M_{\odot}$), about 100 times less massive than the known LyC leakers at $z \sim 0.3$ \citep{Flury2022a, Flury2022b}. Their UV luminosities range from $M_{\rm UV}\sim-16$ to $-15$.  The high nebular temperatures ($T_e$\,$\sim$\,20,000K) seen in these galaxies further underscore their extreme nature, reflecting their low-metal contents. This unique combination of characteristics (e.g., low metallicity, high ionization, and young starburst ages) positions the CSFGs as critical laboratories for testing and calibrating photoionization models that aim to replicate the conditions of early-universe galaxies, especially their critical resemblance to the faintest star-forming galaxies and/or the extreme galaxies in the EoR \citep{Fujimoto2023a, Tang2023a, Atek2024a, Topping2024b}.
 
\subsection{UV Spectroscopy}
The rest-frame UV spectra ($1200$\,--\,$2000$\AA) of our five CSFGs were obtained using HST's Cosmic Origins Spectrograph (COS) and Space Telescope Imaging Spectrograph (STIS) instruments to capture emission lines, including \ion{C}{4} $\lambda$1550, \ion{He}{2} $\lambda$1640, \ion{O}{3}] $\lambda$1663, and \ion{C}{3}] $\lambda$1908 (Program ID 16213, PI: S. Ravindranath). We used the FUV G140L grating for COS with a typical spectral resolution of $R\sim3000$ and the NUV-MAMA G230L grating for STIS observations, which provides a spectral resolution of $R\sim700$ at the wavelengths of \ion{C}{3}] emission lines of our targets. The exposure times vary depending on the brightness of galaxies and the predicted emission line fluxes. We summarize the HST observations and the targets' properties in Table \ref{tab:targets}. 

\begin{figure}[t]
\centering
\includegraphics[width=\columnwidth]{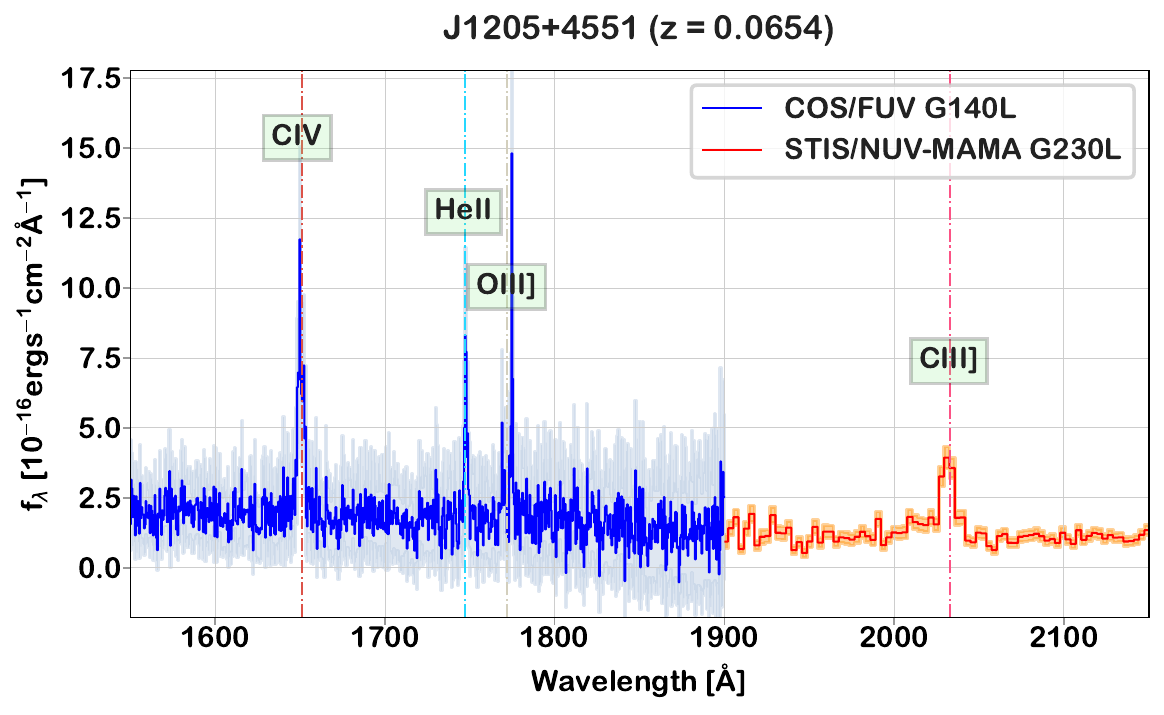}
\caption{We showcase the COS (blue) and STIS (red) spectra of one of our targets: J1205+4551. The \ion{C}{4} $\lambda$1550, \ion{He}{2} $\lambda$1640, \ion{O}{3}] $\lambda$$\lambda$1661,1666, and \ion{C}{3}] $\lambda$1908 emission lines are clearly detected, highlighting the intense \ion{C}{4} emission line.} 
\label{fig:J1205_spectra}
\end{figure}

We downloaded the Hubble Advanced Spectral Products (HASP)\footnote{https://archive.stsci.edu/missions-and-data/hst/hasp} for our COS and STIS observations (\texttt{\_cspec.fits} files). The HASP program provides high-quality coadded 1D spectra for COS and STIS observations based on an automated coaddition approach. We separately followed coadd steps to obtain custom coadd products from individual observations (\texttt{\_x1d.fits} files) and found no significant improvement in the quality of final coadd 1D spectra compared to the HASP products of our observations. We present the HASP COS and STIS spectra of J1205+4551 with key UV emission-line features in Figure \ref{fig:J1205_spectra}.

\begin{figure*}[t]
\centering
\includegraphics[width=1.0\textwidth]{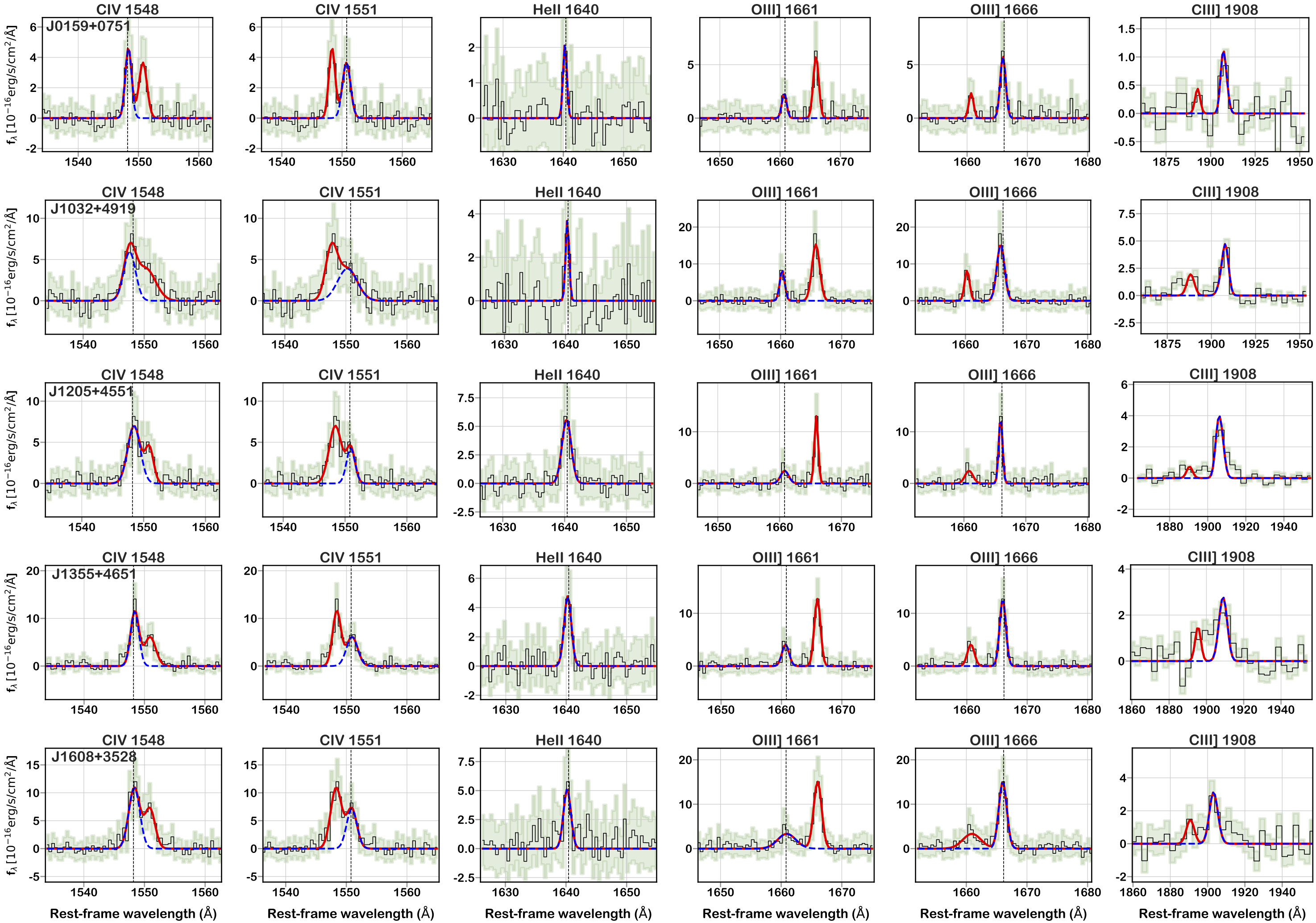}
\caption{Rest-UV emission lines detected in the COS and STIS spectra of our targets. The five left columns show emission lines discovered in COS spectra, while \ion{C}{3}] emission lines in the last column are found in STIS observations. The observed HST spectra are presented as black histograms with uncertainties as shaded. The blue dashed curves show individual emission line fits as labeled, while the red solid curves represent the continuum-subtracted multi-emission-line fits. \ion{C}{4} doublets are marginally resolved with double Gaussian models. While \ion{C}{3}]1908 doublets are blended and barely resolved in our low-resolution STIS spectra, we fit \ion{C}{3}]1908 with combined double Gaussians with fixed line separations to capture the broadened line shapes.}
\label{fig:elines}
\end{figure*}

\subsection{Optical Spectroscopy}
The optical spectra of our CSFGs targets were previously analyzed in \cite{Izotov2017a} using LBT observations. In addition, \cite{Izotov2020a} analyzed the SDSS spectra \citep[Data Release 14;][]{Abolfathi2018a} of four of our targets: J0159+0751, J1032+4919, J1205+4551, and J1355+4651, but excluding J1608+3528. For our study, we obtained the latest SDSS optical spectra from the SDSS Data Release 18 \citep[DR18;][]{Kollmeier2019a, Almeida2023a} and measured the emission line quantities separately, as same as done for UV emission lines in Section 3.1.  While our derived quantities are mostly consistent with previous studies, we rely on our measurements to maintain consistency in our analyses.

\begin{deluxetable*}{cccccc}
\label{tab:eline}
\tabletypesize{\footnotesize}
\tablecaption{Rest-UV Emission-Line Properties from HST Spectra} 
\tablehead{
\colhead{Line/Line-Ratio} & \colhead{J0159+0751} & \colhead{J1032+4919}  & \colhead{J1205+4551} & \colhead{J1355+4651} & \colhead{J1608+3528}
}
\startdata
\multicolumn{6}{c}{Dust-Corrected Emission-Line Fluxes ($\times10^{-16}$\,egs\,s$^{-1}$\,cm$^{-2}$)} \\
{\ion{C}{4} $\lambda$1548} & {$63.3\pm19.9$}    & {$29.3\pm14.3$}& {$19.2\pm4.9$} & {$20.3\pm6.6$}  & {$28.1\pm9.7$}\\		
{\ion{C}{4} $\lambda$1551} & {$63.7\pm19.6$}    & {$30.5\pm12.5$}& {$7.9\pm4.2$} & {$14.7\pm6.4$}  & {$17.1\pm10.1$}\\
{\ion{He}{2} $\lambda$1640} & {$20.3\pm14.3$}   & {$7.2\pm5.3$} & {$9.8\pm2.9$}  & {$7.5\pm2.4$}  & {$13.2\pm4.9$} \\
{\ion{O}{3}] $\lambda$1661} & {$22.1\pm16.0$}   & {$16.0\pm8.4$} & {$3.9\pm2.6$}  & {$6.2\pm3.2$}  & {$20.5\pm10.4$} \\
{\ion{O}{3}] $\lambda$1666} & {$62.6\pm18.7$}   & {$45.9\pm10.9$} & {$10.6\pm2.8$}  & {$17.0\pm3.5$}  & {$41.7\pm9.3$} \\
{\ion{C}{3}] $\lambda$1908$^{\dagger}$} & {$32.4\pm6.3$}    & {$43.3\pm5.6$} & {$18.8\pm2.6$}  & {$15.4\pm1.6$}  & {$17.8\pm3.4$} \\
\hline
\multicolumn{6}{c}{Rest-Frame Equivalent Width (\AA)} \\
{\ion{C}{4} $\lambda$1550$^{\dagger}$}  & {$13.9\pm3.1$} & {$11.4\pm3.6$} & {$14.1\pm3.3$} & {$16.0\pm4.2$} & {$15.4\pm4.8$}\\
{\ion{He}{2} $\lambda$1640}             & {$2.8\pm2.0$} & {$1.5\pm1.1$} & {$5.4\pm1.6$} & {$3.8\pm1.2$} & {$3.2\pm1.3$}\\
{\ion{O}{3}] $\lambda$1663$^{\dagger}$} & {$12.7\pm3.7$} & {$13.8\pm3.1$} & {$8.2\pm2.1$} & {$12.2\pm2.5$} & {$15.6\pm3.5$}\\
{\ion{C}{3}] $\lambda$1908$^{\dagger}$} & {$11.3\pm2.2$} & {$15.3\pm2.0$} & {$16.9\pm2.3$} & {$12.7\pm1.3$} & {$10.9\pm2.1$}\\
\hline
\multicolumn{6}{c}{Emission-Line Ratios$^{\dagger\dagger}$} \\
{\ion{C}{4}/\ion{He}{2}}   & {$6.3\pm4.6$} & {$8.3\pm6.7$} & {$2.8\pm1.0$} & {$4.7\pm2.0$} & {$3.4\pm1.7$}\\
{\ion{O}{3}]/\ion{He}{2}}  & {$4.2\pm3.2$} & {$8.6\pm6.6$} & {$1.5\pm0.6$} & {$3.1\pm1.2$} & {$4.7\pm2.0$}\\
{\ion{C}{3}]/\ion{He}{2}}  & {$1.6\pm1.2$} & {$6.0\pm4.5$} & {$1.9\pm0.6$} & {$2.1\pm0.7$} & {$1.3\pm0.6$}\\
{\ion{C}{4}/\ion{C}{3}}    & {$3.9\pm1.1$} & {$0.7\pm0.5$} & {$1.4\pm0.4$} & {$2.3\pm0.6$} & {$2.5\pm0.9$}\\
\enddata
\tablenotetext{}{
$^{\dagger}$\footnotesize Represent the combined doublet lines.\\
$^{\dagger\dagger}$ For the doublet emission lines, doublet-combined line fluxes are used to measure the line ratios.
}
\end{deluxetable*}

\section{Analysis}
\subsection{Emission-Line Measurements}
As our COS and STIS observations detect continuum fluxes in addition to emission lines, we fit emission lines together with a continuum simultaneously. We model a fitting function that combines a polynomial continuum model and multiple Gaussian models for emission lines. The Gaussian models are centered at the expected wavelengths of UV emission lines.  Table \ref{tab:eline} lists the key emission lines. \ion{C}{4}, \ion{He}{2}, and \ion{O}{3}] lines are covered in the COS observations while \ion{C}{3}] is observed with STIS. 

The COS and STIS spectra were first corrected for the Galactic extinction. The line-of-sight dust reddening ranges from \textit{E(B-V)} = 0.0064 to 0.0744 \citep{Schlafly2011a}\footnote{https://irsa.ipac.caltech.edu/applications/DUST/} based on the \cite{Cardelli1989a} reddening law, assuming $R_V=3.1$. Then, we model continuum spectra with polynomial functional forms excluding emission line regions and model emission lines with 1D Gaussian functions. For emission-line models, we fix the line centers at the expected wavelengths of emission lines with given spectroscopic redshifts. We model each of \ion{C}{4} doublets with single Gaussian functions as they are well-fitted with the combined spectra of the two Gaussians. Although \ion{C}{3}] $\lambda$1908 doublets are blended and barely resolved in our low-resolution SITS spectra, we model \ion{C}{3}] emission with combined double Gaussians with fixed line separations to best reproduce the broadened line shapes, while fixing the line widths of the doublet. We present our emission line fitting in Figure \ref{fig:elines}.

Interestingly, the STIS spectra present weak signals near \ion{C}{3}] $\lambda$1908 (see the \ion{C}{3}] $\lambda$ panels in Figure \ref{fig:elines}). We also fit to \ion{Si}{3}] $\lambda$1892, shown as weaker Gaussian peaks left to \ion{C}{3}]. However, there are still additional signals noticed around \ion{C}{3}]. These extreme (low-metallicity, young, high ionization) galaxies may have other emission lines (e.g., [\ion{S}{2}] $\lambda$1902\footnote{From the NIST Atomic Spectra Database\citep{NIST_ASD}}) that may be responsible for the broad appearance of the line. However, we cannot fit the additional lines that may be intrinsically weak and do not have sufficient signal-to-noise to provide reliable measurements. To avoid including such signals to \ion{C}{3}] line measurements, we fix the line widths in Gaussian models same as measured from \ion{O}{3}] $\lambda$1666 in COS spectra. Specifically, we measured the instrumental-broadening-corrected line widths of \ion{O}{3}] $\lambda$1666 lines and convolved them with the STIS spectral resolutions. We measured the physical quantities of the \ion{C}{3}] lines from the combined double Gaussian fitting.

Line fluxes are measured as the area under the best-fit Gaussians. To estimate EWs, we use the modeled continuum spectra to determine the continuum levels at the centers of emission lines. The uncertainties of the key physical quantities are obtained from the bootstrapping approach. We generated 1000 simulated spectra that were randomly drawn using the error spectra, measured key quantities from the synthetic spectra, and adopted standard deviations as 1$\sigma$ uncertainties.

The emission line fluxes are also corrected for the intrinsic dust attenuation based on \textit{E(B-V)} values. We derived the \textit{E(B-V)} values for our sources by measuring the Balmer decrements from their SDSS spectra. Then, we corrected the UV emission line fluxes with the reddening values, using the \cite{Reddy2015a} UV dust attenuation curve with $R_V = 2.505$. Table \ref{tab:eline} includes the dust-corrected fluxes, the rest-frame EWs of the detected UV emission lines, and the key line ratios analyzed in this work.

It is worth discussing that \ion{C}{4} is often dominated by stellar emission by luminous O stars, which is characterized by a broad P-Cygni profile reflecting strong stellar winds \citep{Shapley2003a, Steidel2016a, Smith2023a}.  Our spectra lack P-Cygni absorption from these massive stars with no characteristic features of Wolf-Rayet stars (i.e. broad bumps at $\sim$ 4650\AA\ and 5800\AA) in the observed spectra \citep{Izotov2017a}. Thus, we consider \ion{C}{4} in our sources dominated by nebular components as seen in high-$z$ galaxies \citep[e.g.,][]{Stark2015b, Mainali2017a} or local star-forming dwarfs \citep[e.g.,][]{Berg2016a, Berg2019b, Senchyna2017a, Senchyna2019a, Senchyna2022a, Wofford2021a, Mingozzi2022a}. \ion{He}{2} can also have stellar contributions, characterized in general by its broad line shape compared to other nebular lines.  However, we do not find such broad-line features in our sources within given detection significance.

We measure optical-line properties from SDSS spectra in the same manner. We first correct the optical spectra for the Galactic extinction, and the internal dust attenuations are corrected based on the E(B-V) values, derived from the Balmer decrements, using the \cite{Cardelli1989a} reddening law with $R_V=3.1$. Table \ref{tab:sdss} includes key optical emission line measurements of our sample used in the analysis of this work.

\begin{deluxetable*}{cccccc}
\label{tab:sdss}
\tabletypesize{\footnotesize}
\tablecaption{Rest-Optical Emission-Line Properties from SDSS Spectra} 
\tablehead{
\colhead{Line/Property} & \colhead{J0159+0751} & \colhead{J1032+4919}  & \colhead{J1205+4551} & \colhead{J1355+4651} & \colhead{J1608+3528}
}
\startdata
\multicolumn{6}{c}{Dust-Corrected Emission-Line Fluxes ($\times10^{-16}$\,egs\,s$^{-1}$\,cm$^{-2}$)} \\
{[\ion{O}{2}] $\lambda$3727} & {$25.11\pm1.55$} & {$41.68\pm2.12$} & {$15.52\pm0.63$} & {$6.24\pm0.80$} & {$18.89\pm2.36$}\\
{H$\delta$}  & {$49.4\pm1.0$} & {$72.1\pm1.3$} & {$78.0\pm1.3$} & {$34.9\pm2.4$} & {$45.7\pm1.4$}\\
{H$\gamma$}  & {$110.4\pm1.9$} & {$164.2\pm3.1$} & {$175.6\pm3.9$} & {$76.2\pm3.2$} & {$96.4\pm2.4$}\\
{[\ion{O}{3}] $\lambda$4363} & {$30.63\pm0.56$} & {$33.49\pm0.51$} & {$12.07\pm0.24$} & {$4.02\pm0.28$} & {$18.07\pm0.53$}\\
{\ion{He}{2} $\lambda$4686} & {$1.66\pm0.28$} & {$1.71\pm0.12$} & {$3.15\pm0.16$} & {$<0.20$} & {$1.44\pm0.26$}\\
{H$\beta$}  & {$133.27\pm2.99$$^{*}$} & {$153.26\pm3.59$} & {$93.17\pm1.72$$^{*}$} & {$17.67\pm0.67$} & {$83.49\pm1.42$}\\
{[\ion{O}{3}] $\lambda$4959} & {$261.98\pm7.72$} & {$294.98\pm9.98$$^{*}$} & {$130.68\pm3.71$} & {$36.33\pm0.72$} & {$232.01\pm3.22$}\\
{[\ion{O}{3}] $\lambda$5007} & {$780.69\pm23.01$$^{*}$} & {$879.05\pm29.75$$^{*}$} & {$389.42\pm11.04$$^{*}$} & {$107.65\pm1.86$} & {$695.06\pm8.49$}\\
{\ion{He}{1} $\lambda$5876} & {$12.47\pm0.43$} & {$16.86\pm0.31$} & {$10.71\pm0.18$} & {$5.79\pm0.34$} & {$9.46\pm0.24$}\\
{[\ion{S}{2}] $\lambda$6717} & {$1.32\pm0.18$} & {$3.32\pm0.11$} & {$0.90\pm0.07$} & {$0.85\pm0.08$} & {$1.56\pm0.15$}\\
{[\ion{S}{2}] $\lambda$6731} & {$1.10\pm0.18$} & {$2.82\pm0.10$} & {$0.84\pm0.07$} & {$0.55\pm0.11$} & {$1.56\pm0.15$}\\
\hline
\multicolumn{6}{c}{Rest-Frame Equivalent Width (\AA)} \\
{H$\beta$} &  {$338.8\pm12.6$$^{*}$} & {$452.6\pm11.5$} & {$512.7\pm15.0$$^{*}$} & {$221.8\pm8.4$} & {$259.8\pm4.0$}\\
{[\ion{O}{3}] $\lambda$4959} & {$699.5\pm22.8$$^{*}$} & {$317.1\pm6.1$} & {$726.7\pm19.1$} & {$480.5\pm9.6$} & {$754.2\pm8.9$}\\
{[\ion{O}{3}] $\lambda$5007} & {$2084.7\pm68.0$$^{*}$} & {$944.8\pm18.2$$^{*}$} & {$2165.6\pm57.0$$^{*}$} & {$1458.1\pm27.0$} & {$2310.9\pm28.8$}\\
\hline
{\textit{E(B-V)}$^{\dagger}$} & {$0.29^{+0.11}_{-0.12}$} & {$0.08^{+0.10}_{-0.08}$} & {$0.0$} & {$0.0$} & {$0.01^{+0.08}_{-0.01}$}\\
\enddata
\tablenotetext{}{
$^{*}$\footnotesize Lines in SDSS spectra are clipped or anomalous. The values are adopted as $2.132\times$H$\gamma$ for H$\beta$ and $2.98\times$[\ion{O}{3}] $\lambda$4959 for [\ion{O}{3}] $\lambda$5007. For J1032+4919, [\ion{O}{3}] $\lambda$4959 is also anomalous, off-Gaussian in the SDSS DR 18 spectrum. Thus, we adopt the [\ion{O}{3}] $\lambda$4959 line flux from \cite{Izotov2020a}, which was measured from the SDSS DR 14 spectrum, by applying a scaling factor based on the H$\beta$ flux ratio between \cite{Izotov2020a} and this work.\\
$^{\dagger}$ Dust reddening values are estimated from the Balmer decrements of H$\beta$/H$\gamma$ (or H$\gamma$/H$\delta$ when H$\beta$ is anomalous).
}
\end{deluxetable*}

\subsection{Temperature, Density and Oxygen Abundance} 
We derive electron temperatures ($T_e$), number densities ($n_e$), and oxygen abundances (12+log(O/H)) based on SDSS optical emission line properties (Table \ref{tab:sdss}). We derive $T_e$ ([\ion{O}{3}]) based on the flux ratio of [\ion{O}{3}] ($\lambda$4959+$\lambda$5007)/$\lambda$4363, following Eq. (4) of \cite{Nicholls2020a}, and use the scaling relation from \cite{Arellano-Cordova2020a} to derive $T_e$ ([\ion{O}{2}]). The electron number density is derived from the [\ion{S}{2}] ratio using {\sc PyNeb} \citep{Luridiana2015a}.  We derive 12+log(O/H) from the \cite{Peng2023a} calibrations for O$^{2+}$/H$^{+}$ and O$^{+}$/H$^{+}$. Although we do not detect \ion{O}{4}] in our UV spectra, with the fact that our sources present high-ionization lines including \ion{He}{2} (the ionization energy comparable to \ion{O}{4}]), we examine O$^{3+}$/H$^{+}$ contributions of the overall oxygen abundance \citep{Berg2018a}. We estimate O$^{3+}$/H$^{+}$ following the approximated description in \cite{Izotov2006a} (Eq. 17), using the \ion{He}{2} $\lambda$4686 and \ion{He}{1} $\lambda$5876 line fluxes in the SDSS spectra of our sources. To calculate ionic abundances, we use the \texttt{getIonAbundance} task in {\sc PyNeb} \citep{Luridiana2015a}. We include O$^{3+}$/H$^{+}$ in our oxygen abundance measurements, although their contributions to the total oxygen abundances are still minimal ($\lesssim$1\%).

We note that the electron densities measured from more consistent ionization zones (i.e., \ion{C}{3}]) are more appropriate for high-ionization lines. Particularly, \cite{Mingozzi2022a} find electron densities estimated from \ion{C}{3}] overall $\sim2$ dex higher than the ones obtained from ([\ion{S}{2}]).  However, we cannot resolve \ion{C}{3}] doublets in our spectra; thus, we adopt the electron density measured from [\ion{S}{2}]. Our targets have high electron temperatures and densities ranging from $T_e $([\ion{O}{3}]) $=17420$ to $22260$\,K and from $n_e $([\ion{S}{2}]) $=210$ to $880$\,cm$^{-3}$, respectively. They are low in metal contents with 12+log(O/H) $\leq7.8$. We list temperatures, densities, and oxygen abundances of our sources in Table \ref{tab:CO}.

\subsection{C/O Abundance}
Typically, relative C/O abundance can be measured using the C$^{2+}$/O$^{2+}$ ionic abundance ratio, applying a carbon ionization correction factor (ICF) to mainly account for the contribution of C$^{3+}$ ions \citep[e.g.,][]{Berg2019a, Jones2023a, Hu2024a, Topping2024a}. First, we follow the same approach to draw the C/O abundances of our sample. We assume the C$^{2+}$ zone has the same temperature and electron density as the O$^{2+}$ zone.  We use the photoionization model-derived ICF available from \citet{Berg2019a} to correct the contributions from other C ions (mainly C$^{3+}$). \cite{Berg2019a} used {\sc CLOUDY} \citep{Ferland2013a} and BPASS \citep{Eldridge2016a, Stanway2016a} to estimate the ICF as a function of ionization parameter. We use the relations in \cite{Berg2019a} to infer the interpolated ionization parameters (log$U$) and carbon ICFs with O$_{32}$ ratios and metallicities of our sources. The derived ionization parameters span log$U=[-1.77\pm0.03$, $-1.33\pm0.06$], and we obtain the ICFs from $1.14\pm0.02$ to $1.33\pm0.04$. We calculate C$^{2+}$ and O$^{2+}$ ionic abundances using the dust-corrected line flux ratios, \ion{C}{3}] $\lambda$1908/\ion{O}{3}] $\lambda$1666. We exclude \ion{O}{3}] $\lambda$1661 as their detection significance is low at a 1--2$\sigma$ level \citep[e.g.,][]{Berg2016a, Berg2019a}. The resulting C/O abundance ratios are from log(C/O)$_{\rm ICF}$ $=-1.35\pm0.13$ to $-0.74\pm0.14$. 

Alternatively, rather than applying a carbon ICF, a C/O abundance can be calculated by considering C$^{3+}$ contributions directly from \ion{C}{4}/\ion{C}{3}] (for C$^{3+}$/C$^{2+}$), while ignoring C${^+}$ contribution based on high ionization conditions of our sources \citep{Curti2024a}. We follow \cite{Curti2024a} to calculate the total C/O abundances, which account for O$^{3+}$ contributions to the total oxygen abundance. We first calculate C$^{2+}$/H$^{+}$ and C$^{3+}$/H$^{+}$ abundances as: 
\begin{equation}    
\frac{\rm C^{2+}}{\rm H^+} = \frac{\rm C^{2+}}{\rm O^{2+}} \times \frac{\rm O^{2+}}{\rm H^+}
\end{equation}
\begin{equation}    
\frac{\rm C^{3+}}{\rm H^+} = \frac{\rm C^{3+}}{\rm O^{2+}} \times \frac{\rm O^{2+}}{\rm H^+}.
\end{equation}
Then, we set the total carbon abundance (C/H) as:
\begin{equation}    
\frac{\rm C}{\rm H} = \frac{\rm C^{2+}}{\rm H^+} + \frac{\rm C^{3+}}{\rm H^+},
\end{equation}
where we consider C$^{+}$ contribution negligible. 

Lastly, we obtain the total C/O abundances as: 
\begin{equation}    
\frac{\rm C}{\rm O} = \left({\frac{\rm C}{\rm H}}\right) \Big/ \left({\frac{\rm O}{\rm H}} \right).
\end{equation} 
The estimated C/O abundances span from log(C/O) $=-1.12\pm0.14$ to $-0.61\pm0.15$.

We also derive log(C/O) using the empirical calibration given in \cite{Perez-Montero2017a}, which is based on the line flux ratio C3O3 ($\equiv$ (\ion{C}{3}]$\lambda$1908+\ion{C}{4}$\lambda$1550)/\ion{O}{3}]$\lambda$1663). This approach provides C/O abundances, ranging from log(C/O)$_{\rm PM}=-1.06\pm0.13$ to $-0.67\pm0.12$. We summarize the physical quantities of individual sources regarding C/O abundance estimates in Table \ref{tab:CO}.

While the C/O measurements are consistent between the different approaches within their uncertainties, the values based on the carbon ICF (log(C/O)$_{\rm ICF}$) from the photoionization models in \cite{Berg2019a} provide consistently lower C/O abundances than the others. This comparison suggests that the carbon ICFs based on O$_{32}$-based ionization parameters, following the \cite{Berg2019a} description, could underestimate the contributions from C$^{3+}$ for extreme \ion{C}{4} emitters. Being benefited from the \ion{C}{4}/\ion{C}{3}] flux ratios available, we consider the log(C/O) values obtained via these direct estimates of the C$^{3+}$/C$^{2+}$ relative abundance as fiducial C/O measurements.

\begin{deluxetable*}{lccccc}
\label{tab:CO}
\tabletypesize{\footnotesize}
\tablecaption{Electron Temperatures, Electron Number Densities, and Element Abundances}
\tablehead{
\colhead{Property\qquad } & \colhead{\qquad J0159+0751\qquad } & \colhead{\qquad J1032+4919\qquad }  & \colhead{\qquad J1205+4551\qquad } & \colhead{\qquad J1355+4651\qquad } & \colhead{\qquad J1608+3528\qquad }
}
\startdata
{ $T_e$ ([\ion{O}{3}])\,/\,K\qquad }        & {\qquad $22260\pm440$\qquad }       & {\qquad $21830\pm450$\qquad }       & {\qquad $19200\pm340$\qquad }       & {\qquad $21520\pm1020$\qquad }      & {\qquad $17420\pm280$\qquad }\\
{ $n_e$ ([\ion{S}{2}])\,/\,cm$^{3}$\qquad } & {\qquad $530^{+670}_{-370}$\qquad } & {\qquad $410^{+140}_{-120}$\qquad } & {\qquad $670^{+410}_{-310}$\qquad } & {\qquad $210^{+330}_{-160}$\qquad } & {\qquad $880^{+610}_{-410}$\qquad }\\
{ 12+log(O/H)\qquad }                       & {\qquad $7.44\pm0.02$\qquad }       & {\qquad $7.45\pm0.02$\qquad }       & {\qquad $7.43\pm0.02$\qquad }       & {\qquad $7.49\pm0.04$\qquad }       & {\qquad $7.81\pm0.02$\qquad }\\
{ [\ion{O}{3}]/[\ion{O}{2}]\qquad }         & {\qquad $41.52\pm2.74$\qquad }      & {\qquad $28.17\pm1.62$\qquad }      & {\qquad $33.52\pm1.56$\qquad }      & {\qquad $23.08\pm2.98$\qquad }      & {\qquad $49.07\pm6.15$\qquad }\\
{ log$U$\qquad }                            & {\qquad $-1.51\pm0.06$\qquad }      & {\qquad $-1.69\pm0.03$\qquad }      & {\qquad $-1.61\pm0.02$\qquad }      & {\qquad $-1.77\pm0.05$\qquad }      & {\qquad $-1.33\pm0.06$\qquad }\\
{ log(C$^{2+}$/O$^{2+}$)\qquad }            & {\qquad $-1.32\pm0.17$\qquad }      & {\qquad $-1.07\pm0.14$\qquad }      & {\qquad $-0.83\pm0.14$\qquad }      & {\qquad $-1.09\pm0.11$\qquad }      & {\qquad $-1.48\pm0.13$\qquad }\\
{ log(C$^{3+}$/C$^{2+}$)\qquad }            & {\qquad $0.21\pm0.14$\qquad }       & {\qquad $-0.23\pm0.17$\qquad }      & {\qquad $-0.17\pm0.13$\qquad }      & {\qquad $-0.01\pm0.12$\qquad }      & {\qquad $0.12\pm0.19$\qquad }\\
{ log(C/O) (fiducial)\qquad }               & {\qquad $-0.91\pm0.16$\qquad }      & {\qquad $-0.89\pm0.14$\qquad }      & {\qquad $-0.61\pm0.15$\qquad }      & {\qquad $-0.81\pm0.13$\qquad }      & {\qquad $-1.12\pm0.14$\qquad }\\
{ ICF (C)\qquad }                           & {\qquad $1.27\pm0.03$\qquad }       & {\qquad $1.18\pm0.01$\qquad }       & {\qquad $1.22\pm0.01$\qquad }       & {$1.14\pm0.02$\qquad }              & {\qquad $1.33\pm0.04$\qquad }\\
{ log(C/O)$_{\rm ICF}$\qquad }              & {\qquad $-1.22\pm0.17$\qquad }      & {\qquad $-1.00\pm0.14$\qquad }      & {\qquad $-0.74\pm0.14$\qquad }      & {\qquad $-1.03\pm0.11$\qquad }      & {\qquad $-1.35\pm0.13$\qquad }\\
{ log(C/O)$_{\rm PM^{\dagger}}$\qquad }     & {\qquad $-0.85\pm0.13$\qquad }      & {\qquad $-0.89\pm0.11$\qquad }      & {\qquad $-0.67\pm0.12$\qquad }      & {\qquad $-0.80\pm0.12$\qquad }      & {\qquad $-1.06\pm0.13$\qquad }\\
\enddata
\tablenotetext{}{
$^{\dagger}$\footnotesize Measurements based on the empirical calibration in \cite{Perez-Montero2017a} based on the line flux ratio \ion{C}{3}]$\lambda$1908+\ion{C}{4}$\lambda$1550)/\ion{O}{3}]$\lambda$1663.
}
\end{deluxetable*}

\begin{figure*}[th]
\centering
\includegraphics[width=\textwidth]{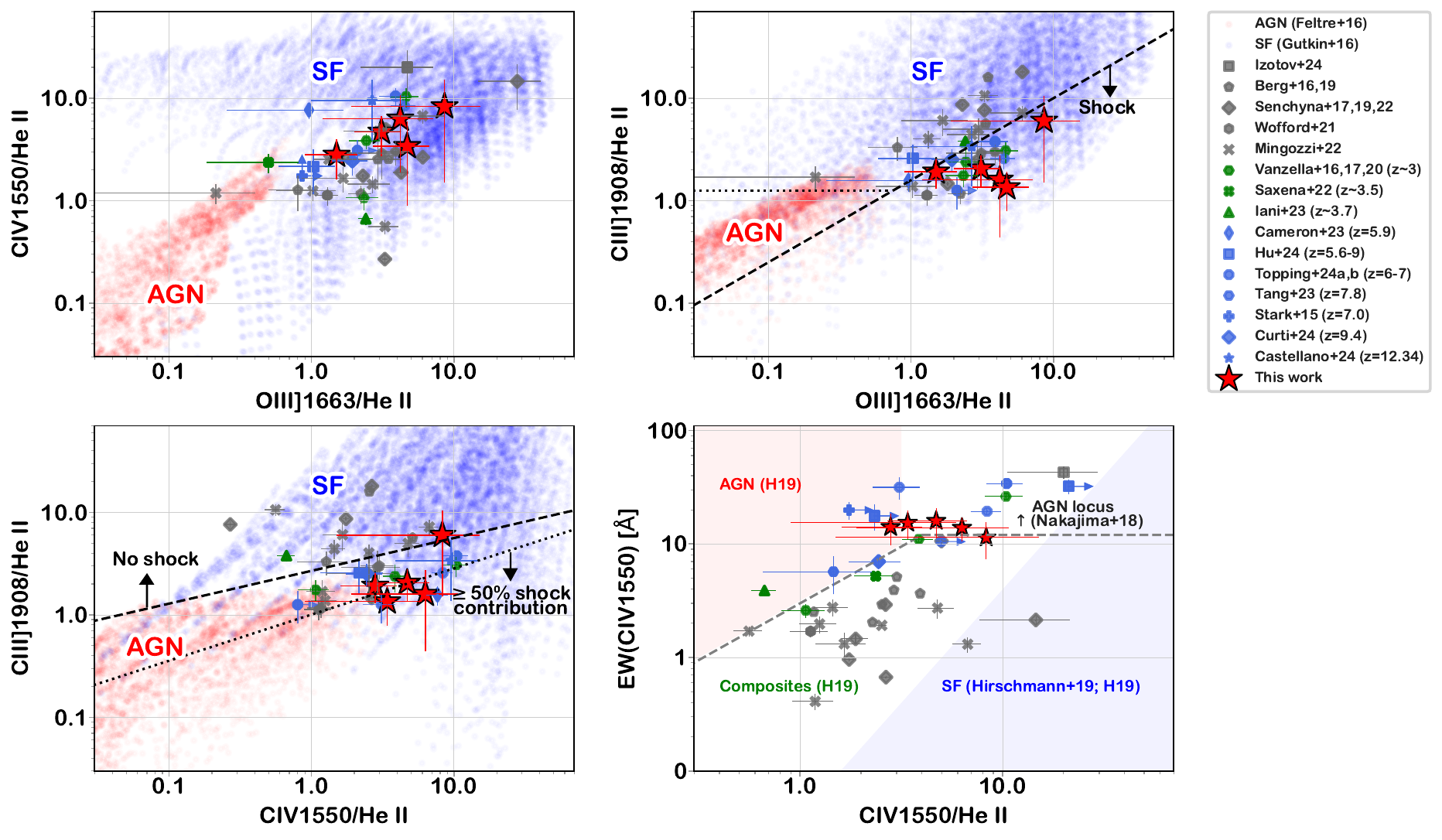}
\caption{Rest-UV diagnostic diagrams. \textit{Top left}: \ion{C}{4} $\lambda$1550/\ion{He}{2} \textit{vs.} \ion{O}{3} 1663/\ion{He}{2} ratios of \ion{C}{4} emitters. \textit{Top right}: \ion{C}{3}] $\lambda$1908/\ion{He}{2} \textit{vs.} \ion{O}{3} 1663/\ion{He}{2}. \textit{Bottom left}: \ion{C}{3}] $\lambda$1908/\ion{He}{2} \textit{vs.} \ion{C}{4} $\lambda$1550/\ion{He}{2}. The blue and red shaded regions represent star-forming galaxies (SF) and active galactic nuclei (AGN), respectively, based on the models by \cite{Gutkin2016a} and \cite{Feltre2016a}. The shock separators are from \cite{Mingozzi2024a} in the top right panel and \cite{Jaskot2016a} in the bottom left panel. \textit{Bottom right}: EW(\ion{C}{4} 1550) \textit{vs.} \ion{C}{4}/\ion{He}{2}. The blue dashed line represents the AGN locus based on the photoionization models in \cite{Nakajima2018a}. Red and blue shades represent AGN and SF regions based on \cite{Hirschmann2019a} with the composite region in between. Gray symbols represent the local samples \citep{Berg2016a, Berg2019a, Senchyna2017a, Senchyna2019a, Senchyna2022a, Wofford2021a, Mingozzi2022a, Izotov2024a}, while $z\sim3$ galaxies are shown as green \citep{Vanzella2016a, Vanzella2017a, Vanzella2020b, Saxena2022a, Iani2023a} and $z\gtrsim6$ galaxies are in blue \citep{Stark2015a, Cameron2024a, Tang2023a, Hu2024a, Topping2024a, Topping2024b, Castellano2024a, Curti2024a}.  Our CSFGs are shown as red stars. These diagnostics suggest no clues for AGN origins in our targets while suggesting possible shock contributions. Our CSFGs preferentially lie in the region occupied by $z\gtrsim6$ galaxies in these diagnostic plots, highlighting that the ionizing conditions in our CSFGs are similar to those in reionization-era \ion{C}{4}-emitting galaxies.} 
\label{fig:C4He2_O3He2}
\end{figure*}

\section{Discussion}
JWST observations have revealed extreme properties of reionization-epoch galaxies, such as high ionizing photon efficiencies and high ionization state of the ISM, presenting prominent emission-line features (e.g., \ion{N}{4}], \ion{C}{4}, and \ion{He}{2}) in their rest-UV spectra \citep[e.g.,][]{Bunker2023a, Tang2023a, Larson2023a, Jung2024a, Witstok2024a, Castellano2024a, Kumari2024a, Curti2024a, Topping2024b}. With their intense high-ionization emission lines, indicative of strong ionizing radiation, these galaxies may play a crucial role in reionization. We have identified local CSFGs with similarly intense \ion{C}{4} emission. In this section, we will discuss the characteristics of our targets in detail.

\subsection{Ionization Sources}
High-ionization lines of reionization-era galaxies observed with JWST spectroscopy (e.g., \ion{C}{4} ann \ion{He}{2}) require sources of high-energy radiation fields that are not well reproduced with standard stellar populations alone. Rather, it may require alternative harder ionizing sources, including AGNs, Very Massive Stars, X-ray binaries, and shocks \citep[e.g.,][]{Shirazi2012a, Olivier2022a, Mingozzi2022a, Schaerer2024a}.  We examine ionizing sources of our local analogs along with reionization-era galaxies.

The \ion{C}{4} $\lambda$1550/\ion{He}{2} \textit{vs.} \ion{O}{3} $\lambda$1663/\ion{He}{2} ratio ``C4He2-O3He2" hereafter) is the commonly used UV diagnostics of \ion{C}{4} emitters to distinguish ionizing sources between star formation and AGN activity \citep[e.g.,][]{Mainali2017a, Senchyna2017a, Izotov2024a}.  We compare the C4He2-O3He2 ratios of our targets with the model predictions of star-forming galaxies \citep[SF;][]{Gutkin2016a} and \citep[AGN;][]{Feltre2016a} in Figure \ref{fig:C4He2_O3He2} (top left). The plot shows a clear separation between the regions occupied by AGNs (red) and star-forming galaxies (blue). Our high-O$_{32}$ CSFGs are located in the SF region with most of \ion{C}{4} emitters in literature, indicating that their ionizing sources are likely young, massive stars rather than AGN activities. 

We additionally explore other UV diagnostics, which, in particular, can also test shock origins. In the top right panel, we show the \ion{C}{3} $\lambda$1908/\ion{He}{2} \textit{vs.} \ion{O}{3} $\lambda$1663/\ion{He}{2} ratio (C3He2-O3He2). \cite{Mingozzi2024a} test this diagnostic with SF, BPASS, AGN, and Shock models, drawing lines to separate AGN and Shock from SF models \citep{Gutkin2016a, Xiao2018a, Feltre2016a, Alarie2019a}. This is shown as dashed and dotted lines: shock and AGN separators from star formation, respectively. Interestingly, our CSFG targets and a majority of reionization-era \ion{C}{4} emitters are classified as having shock contributions. 

The bottom left panel presents the \ion{C}{3}] $\lambda$1908/\ion{He}{2} \textit{vs.} \ion{C}{4} $\lambda$1550/\ion{He}{2} diagnostic (C3He2-C4He2), which includes the \cite{Jaskot2016a} shock models. The dashed and dotted lines separate no shock contributions and $\geq$50\% shock contributions. Although the C3He2-C4He2 diagnostic does not clearly separate AGNs from SF compared to the other diagnostics, our targets barely overlap with AGN models. We highlight that four of our targets are suggested as having significant ($\sim$50\%) shock contributions, with the exception of J1032+4919. J1032+4919 is somewhat distinctive from the other four galaxies in our sample in its Ly$\alpha$ profile; it shows a weak Ly$\alpha$ emission at the bottom of a broad Ly$\alpha$ absorption while the others show strong Ly$\alpha$ emission without absorption features. We will discuss this further in Section 4.3 related to LyC escape.

Lastly, in the bottom right panel, we compares EW(\ion{C}{4} 1550) with the \ion{C}{4}/\ion{He}{2} line ratio (EW(\ion{C}{4})-C4He2). This plot is also used to diagnose the ionizing sources of these galaxies, distinguishing between AGN activity (above the dashed line) and star formation (below the dashed line), based on photoionization models from \cite{Nakajima2018a}. The blue and red shaded regions represent the diagnostics of star formation and AGNs from the \cite{Hirschmann2019a} models, which classify the majority of \ion{C}{4} emitters as composites (in the between SF and AGNs). The presence of some high-redshift galaxies in the AGN region or the composites (in C3He2-C4He2 and EW(\ion{C}{4}-C4He2) indicates that there can be an overlap between the properties of high-ionization star-forming regions and AGNs, responsible for intense \ion{C}{4} emission. Alternatively, high C/O abundance can increase EW(\ion{C}{4} 1550) values. However, our targets show no evidence of increased C/O ratios as measured in Section 3.3, the same as discussed in the recent JWST studies of reionization-era galaxies (see discussion in Section 4.4). Instead, such intense \ion{C}{4} emission is more likely to be observed in young, dust-free sources, thus unattenuated through the resonant scattering of \ion{C}{4} photons. Otherwise, \ion{C}{4} emission can be weakened if internal dust is present. Such dust-free sources with \ion{C}{4} emission were previously observed of $z=3-5$ low-mass young galaxies and super star clusters \citep{Vanzella2016a, Vanzella2017a}. It appears not difficult to find low-dust or nearly dust-free sources with \ion{C}{4} from the EoR \citep[e.g.,][]{Topping2024a}.

In summary, we highlight that compared to other local analogs; our CSFG targets preferentially lie in the region occupied by high-redshift galaxies in these diagnostic plots. This alignment suggests that the ionizing conditions in our CSFGs are similar to those in high-redshift \ion{C}{4}-emitting galaxies. Particularly, the diagnostics containing shock models suggest that shocks (e.g., supernovae explosions, stellar winds, and outflows for either starbursts or AGNs) could be consistent with the emission-line ratios that we see in these galaxies. However, these models do not thoroughly test other possibilities, including e.g., catastrophic cooling and Very Massive Stars \citep{Gray2019a, Danehkar2021a, Schaerer2024a, Carr2024a}. Thus, it is necessary to implement model predictions of other possible extreme sources in future studies to confirm the nature of their ionizing sources.

\begin{figure*}[th]
\centering
\includegraphics[width=\textwidth]{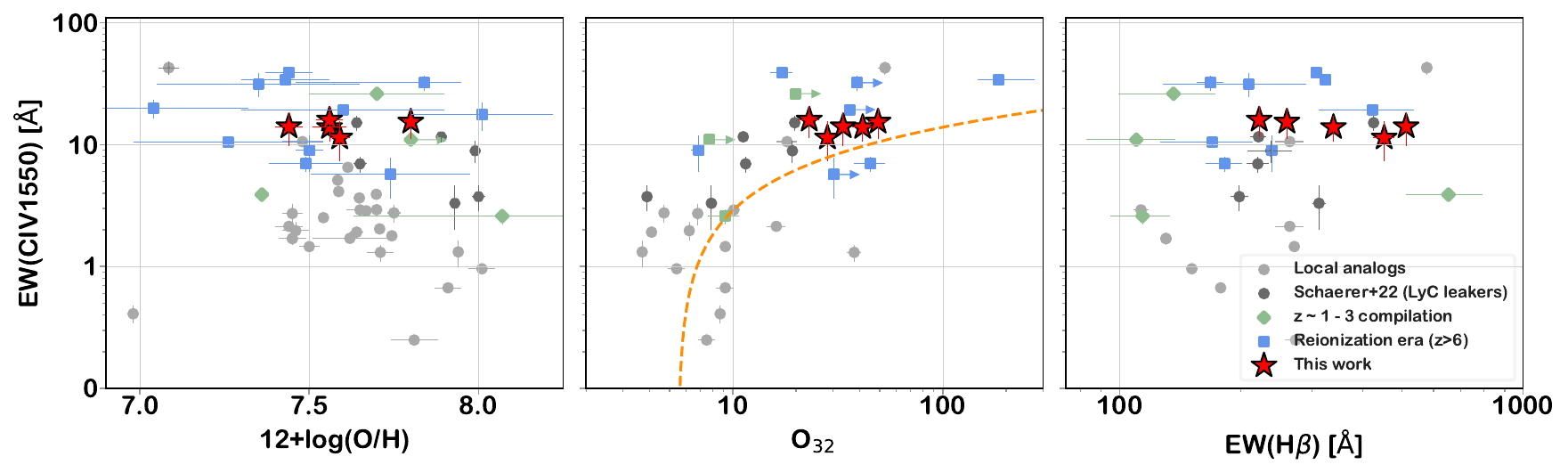}
\caption{\ion{C}{4} EWs \textit{vs.} various galaxy properties. We compare our sources with other \ion{C}{4} emitters in the literature that include LyC leakers. The red stars represent our sample. The local samples are shown as gray symbols \citep{Berg2019a, Senchyna2017a, Senchyna2019a, Senchyna2022a, Wofford2021a, Mingozzi2022a, Schaerer2022a, Izotov2024a}. Local LyC-leaking galaxies are highlighted as dark grays \citep{Schaerer2022a}.  $z\sim1$\,--\,$3$ galaxies are marked in green \citep{Saxena2022a} and, the blue squares represent reionization-era galaxies \citep{Stark2015a, Witstok2021a, Witstok2024a, Tang2023a, Topping2024a, Curti2024a}. 
\textit{Left}: EW(\ion{C}{4} 1550) \textit{vs.} 12+log(O/H). While all \ion{C}{4} emitters present low metallicity (12+log(O/H) $\lesssim8$), our sources are among the highest EW(\ion{C}{4}) emitters at any given metallicity, aligned with reionization-era galaxies.
\textit{Middle}: EW(\ion{C}{4} 1550) \textit{vs.} O$_{32}$. A strong correlation indicates that high EW(\ion{C}{4}) is associated with higher ionization parameters.
\textit{Right}: EW(\ion{C}{4} 1550) \textit{vs.} EW(H$\beta$). All \ion{C}{4} emitters are strong H$\beta$ emitters, with our targets showing the highest EWs, similar to high-redshift \ion{C}{4} emitters and the known LyC leakers among local dwarf \ion{C}{4} emitters. Such high H$\beta$ EWs ($>$ 200\AA) suggest that they are young, intense star-formers.
Overall, our sample exhibits high EW(\ion{C}{4} 1550) values, comparable to both local LyC leakers and $z\gtrsim6$ \ion{C}{4} emitters.} 
\label{fig:C4_EW}
\end{figure*}

\begin{figure*}[th]
\centering
\includegraphics[width=\textwidth]{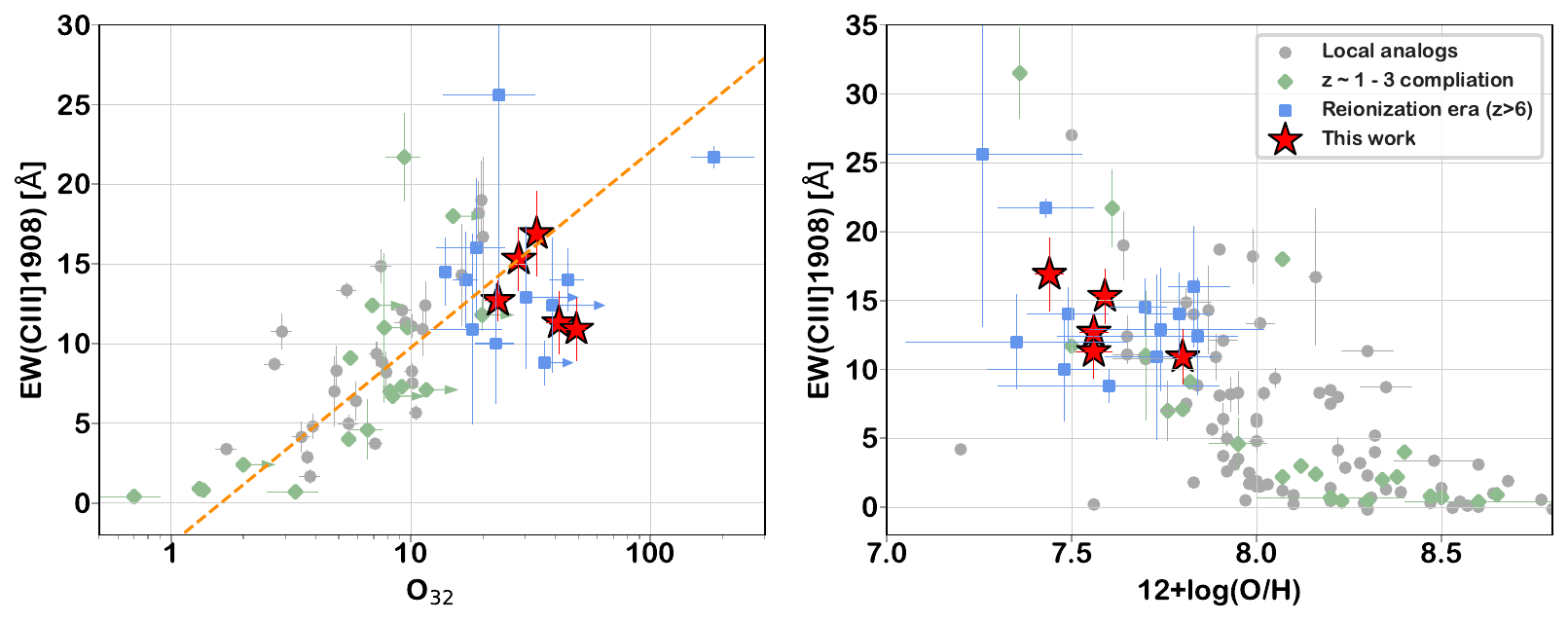}
\caption{The empirical relations of \ion{C}{3}]$ \lambda$1908 EWs \textit{vs.} O$_{32}$ (left) and 12+log(O/H) (right). Our targets are shown as red stars with literature values: local analogs \citep[gray;][]{Giavalisco1996a, Leitherer2011a, Berg2016a, Pena-Guerrero2017a, Senchyna2017a, Senchyna2019a, Ravindranath2020a, Schaerer2022a, Izotov2023a}, $z\sim1$\,--\,$3$ compilation \citep[green;][]{Pettini2000a, Teplitz2000a, Quider2009a, Hainline2009a, Erb2010a, Christensen2012a, Bayliss2014a, James2014a, Stark2014a, Rigby2015a, deBarros2016a, Vanzella2016a, Vanzella2017a, Vanzella2020a, Berg2018a, Mainali2020a, Mainali2023a,  Rigby2021a, Iani2023a}, and high-$z$ galaxies \citep[blue;][]{Tang2023a, Hsiao2023a, Hsiao2024a, Topping2024a, Topping2024b, Castellano2024a, Witstok2024a, Curti2024a}. On the left panel, there is a clear positive correlation with O$_{32}$, indicating that strong \ion{C}{3}] emission requires high ionization conditions in the ISM. The best-fit linear model is presented as the orange dashed line. The right panel presents an anti-correlation trend between \ion{C}{3}] EWs and gas-phase metallicities. Notably, our \ion{C}{4} emitters present intense \ion{C}{3} (EW$>$10\AA) emission with metal-poor conditions, mostly clustered with reionization-era galaxies.}
\label{fig:C3_EW}
\end{figure*}

\subsection{Understanding the Intense Nebular \ion{C}{4} Emission}
Focusing on the physical conditions of our sample in relating them to intense nebular \ion{C}{4}, we compare \ion{C}{4} EWs with various galaxy properties in Figure \ref{fig:C4_EW}. In the left panel, we examine the relationship between EW(\ion{C}{4} 1550) and metallicity, indicated by 12+log(O/H). All the \ion{C}{4} emitters, including our sample, have low metallicity with 12+log(O/H) $\lesssim 8$. The narrow range of metallicities among the \ion{C}{4} emitters prevents detecting a clear correlation between EW(\ion{C}{4}) and metallicity. However, our CSFGs are the highest EW \ion{C}{4} emitters at a given metallicity and tend to lie along with the high redshift galaxies.

The middle panel displays a strong correlation between EW(\ion{C}{4} 1550) and the O$_{32}$ line ratio, a proxy for the ionization parameter; higher O$_{32}$ ratios indicate higher ionization parameters \citep[e.g.,][]{Strom2018a, Kewley2019a, Papovich2022a}. The high O$_{32}$ values ($>20$) in our CSFGs suggest that their intense \ion{C}{4} emission may be primarily linked to their high ionization parameters (log$U$ $>-2$). This correlation highlights the extreme ionization conditions in our sample, comparable to those derived in high-redshift galaxies \citep[log$U\gtrsim-2$ in][]{Tang2023a, Curti2024a, Jung2024a, Topping2024a, Witstok2024a}.  Using the compilation of \ion{C}{4} emitters reported in the literature, we fit a linear function with \ion{C}{4}-detected sources to find an empirical trend of \ion{C}{4} strength with O$_{32}$. We use Orthogonal Distance Regression (ODR) to find a best-fit linear model. The best-fit model is shown as the orange dashed line in the panel. Although there is a considerable scatter, it represents a strong correlation between EW(\ion{C}{4} 1550) and the O$_{32}$ line ratio. The Pearson correlation coefficient between EW(\ion{C}{4} 1550) and logO$_{32}$ is also measured as $r = 0.65$ with $p\sim5\times10^{-5}$, demonstrating their strong correlation. The fitted coefficients of the linear model are listed in Table \ref{tab:fits} with the parameter uncertainties. 

The right panel shows EW(\ion{C}{4} 1550) \textit{vs.} EW(H$\beta$). Our CSFS targets are also strong H$\beta$ emitters ($>$200\AA), suggesting young ($\lesssim$ 3 Myr) stellar populations with intense star formation characterize them \citep[e.g.,][]{Levesque2013a}. Notably, they have the highest EWs of \ion{C}{4} and H$\beta$ among local analogs, sharing the domain with reionization-era \ion{C}{4} emitters. The known local LyC leakers (dark grays) also feature high \ion{C}{4} EWs ($>$3\AA) and H$\beta$ EWs ($\gtrsim$200\AA). It reinforces that our CSFGs and the high-redshift \ion{C}{4} emitters are candidates for strong LyC leakers.

We also examine the relation of \ion{C}{3}] EWs with O$_{32}$, and with 12+log(O/H), in Figure \ref{fig:C3_EW}. In the left panel, there is a clear positive correlation of \ion{C}{3}] EWs with O$_{32}$ with the Pearson correlation coefficient of $r = 0.75$ with $p\sim1\times10^{-10}$. With the highest O$_{32}$ ratios of our sample galaxies as a proxy for highly ionized ISM, these galaxies emit the strongest emission lines, including \ion{C}{3}] as shown here. Most of extreme O$_{32}$ ($>$10) objects including the literature values present strong \ion{C}{3}] emission with EW$>$10\AA. \cite{Tang2023a} discuss strong \ion{C}{3}] from $z>7$ LAEs, being situated at the high end of the \ion{C}{3}] EWs and O$_{32}$ relation. We compiled other \ion{C}{3}] measurements of reionization-era galaxies obtained from the latest JWST observations \citep{Hsiao2023a, Hsiao2024a, Topping2024a, Topping2024b, Castellano2024a, Witstok2024a, Curti2024a}. Our CSFGs are populated at the high end of the linear relation between \ion{C}{3}] and O$_{32}$ (the dashed line), lying together with the reionization-era galaxies.

The anti-correlation of \ion{C}{3}] EWs with gas-phase metallicities have been repeatedly examined in previous studies, although there were sparse \ion{C}{3}] EW measurements available at low metallicities (12+log(O/H) $\lesssim7.5$) which also presents scattered distribution at such low metallicities \citep[e.g.,][]{Jaskot2016a, Ravindranath2020a, Tang2021a, Mainali2023a}.  We place our CSFGs at such low metallicity and high \ion{C}{3}] EWs region with the majority of reionization-era galaxies (the right panel of Figure \ref{fig:C3_EW}). This displays a clearer trend than previous studies, increasing \ion{C}{3}] EWs at lower gas-phase metallicities despite a couple of outliers \citep[I Zw 18 and SBS 1415+437 at the bottom left, showing EWs $<5$\AA\ at 12+log(O/H) $\lesssim7.5$;][]{Leitherer2011a, Pena-Guerrero2017a}.

In summary, our CSFGs have low metallicities (12+log(O/H) $\lesssim7.8$), like other \ion{C}{4} emitters, but exhibit some of the highest \ion{C}{4} EWs ($>$10\AA) at any given metallicities, similarly to reionization-era \ion{C}{4} emitters. A strong correlation is observed between the \ion{C}{4} EWs and the O$_{32}$ line ratio, suggesting that intense \ion{C}{4} emission is primarily associated with high ionization parameters. Particularly, most of high-O$_{32}$-ratio galaxies (O$_{32}>10$) exhibit \ion{C}{4} EWs $> 3$\AA.  The CSFGs are also characterized by the strongest H$\beta$ emission (EW $>$ 200\AA), indicating young starburst ages of $\lesssim$3 Myr. Additionally, these galaxies tend to lie at the high end of the \ion{C}{3}] EW -- O$_{32}$ relation, reinforcing the connection between their high ionization conditions and strongest \ion{C}{3}] emission lines. Collectively, these characteristics align the CSFGs with reionization-era \ion{C}{4}-emitting galaxies, making them excellent analogs for studying conditions conducive to LyC escape of reionization-era galaxies.

\begin{deluxetable}{lcc}
\label{tab:fits}
\tabletypesize{\footnotesize}
\tablecaption{Coefficients for Empirical Trends with [\ion{O}{3}]/[\ion{O}{2}]} 
\tablehead{
\colhead{$y=f(x)$$^{\dagger}$, where $x\equiv$ logO$_{32}$} & \colhead{$c_1$} & \colhead{$c_2$} 
}
\startdata
{EW(\ion{C}{4}): Fig. \ref{fig:C4_EW} (middle)} & {$-8.16\pm3.18$} &{$11.05\pm3.61$}  \\
{EW(\ion{C}{3}]): Fig. \ref{fig:C3_EW} (left)} & {$-2.56\pm1.27$} &{$12.31 \pm1.47$} \\
{\ion{C}{4}/\ion{C}{3}]: Fig. \ref{fig:C43} (right)} & {$-0.59\pm0.19$} &{$1.26\pm0.24$} \\
\enddata
\tablenotetext{}{$^{\dagger}$Coefficients are from linear fits in the form of $y=c_1 + c_2\cdot x$
}
\end{deluxetable}

\begin{deluxetable*}{lccccc}
\label{tab:lya}
\tabletypesize{\footnotesize}
\tablecaption{Ly$\alpha$-Related Properties}
\tablehead{
    \colhead{Property\qquad}  & \colhead{\qquad J0159+0751\qquad}  & \colhead{\qquad J1032+4919\qquad}   & \colhead{\qquad J1205+4551\qquad}  & \colhead{\qquad J1355+4651\qquad} & \colhead{\qquad J1608+3528}
}
\startdata
{ EW(Ly$\alpha$) [\AA]\qquad}                  & {\qquad$160.2\pm2.4$\qquad}    & {\qquad$5.2\pm0.8$\qquad}     &{\qquad$189.5\pm3.2$\qquad}   & {\qquad$175.7\pm1.2$\qquad}   & {\qquad$163$} \\
{$V_{\rm sep}$ [km s$^{-1}$]$^{a}$\qquad}     & {\qquad$311.5\pm4.0$\qquad}    & {\qquad$247.7\pm9.9$\qquad}   & {\qquad$248.8\pm3.2$\qquad}  & {\qquad$259.9\pm4.5$\qquad}   & {\qquad$214\pm30$} \\
{$f_{\rm esc,Ly\alpha}$$^{b}$\qquad}          & {\qquad$0.20\pm0.03$\qquad}    & {\qquad$<$0.01\qquad}         & {\qquad$0.16\pm0.02$\qquad}  & {\qquad$0.39\pm0.05$\qquad}   & {\qquad$0.16\pm0.04$} \\
{$f_{\rm esc,LyC}$$^{c}$\qquad}               & {\qquad$0.09\pm0.03$\qquad}    &{\qquad$-^{\dagger}$\qquad}    &{\qquad$0.19\pm0.03$\qquad}   &{\qquad$0.17\pm0.03$\qquad}    &{\qquad$0.31\pm0.14$} \\
\enddata
\tablenotetext{}{
\footnotesize$^{a}$Velocity peak separation between double peaks in Ly$\alpha$.\\
$^{b}$Ly$\alpha$ escape fraction, a relative ratio of Ly$\alpha$ luminosity to the intrinsic Ly$\alpha$ based on H$\beta$ with the Case B recombination assumption.\\
$^{c}$LyC escape fraction, indirectly inferred from the relation of $V_{\rm sep}$ with $f_{\rm esc,LyC}$ given in Eq. (2) in \cite{Izotov2018b}.\\
$^{\dagger}$The $f_{\rm esc,LyC}$ measurement is not given due to the complex Ly$\alpha$ line profile of J1032+4919; it presents a triple-peaked Ly$\alpha$ emission line at the bottom of a broad Ly$\alpha$ absorption.
}
\tablecomments{All properties are obtained from \cite{Izotov2020a} for J0159+0751, J1032+4919, J1205+4551, and J1355+465. \cite{Jaskot2017a} provides Ly$\alpha$ EW, $V_{\rm sep}$, and $f_{\rm esc,Ly\alpha}$ for J1608+3528. We estimated $f_{\rm esc,LyC}$ for J1608+3528 from its $V_{\rm sep}$ using Eq. (2) in \cite{Izotov2018b}.}
\end{deluxetable*}

\begin{figure*}[th]
\centering
\includegraphics[width=\textwidth]{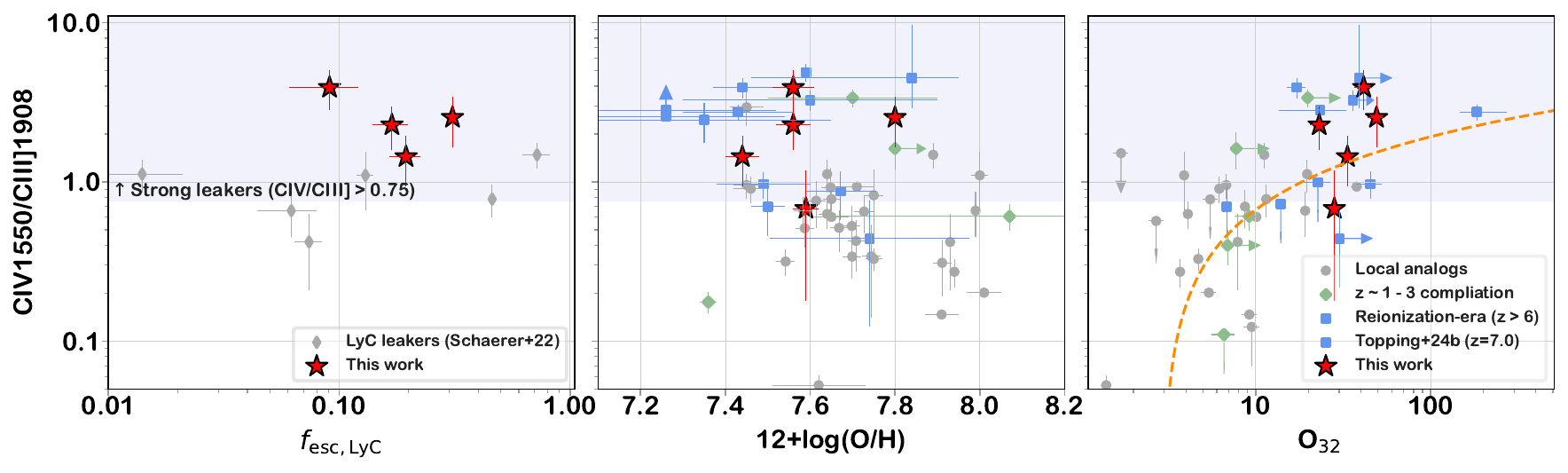}
\caption{\textit{Left}: \ion{C}{4}1550/\ion{C}{3}]1908 (C$_{43}$) \textit{vs.} LyC escape fraction, $f_{\rm esc,LyC}$.  We compare our CSFGs with LyC-leaking \ion{C}{4} emitters \citep{Schaerer2022a}. The region of strong LyC leakers (C$_{43}$$>$0.75), suggested in \cite{Schaerer2022a}, is shaded. The LyC escape fractions of our targets are based on the peak separation of Ly$\alpha$, following Eq.(2) in \cite{Izotov2018b}, suggesting that our CSFGs are candidates of strong LyC leakers.
\textit{Middle}: C$_{43}$ \textit{vs.} 12+log(O/H). Red star symbols represent our targets. High-redshift ($z>6$) samples are shown as blue symbols \citep{Cameron2024a, Tang2023a, Hu2024a, Topping2024a, Topping2024b, Witstok2024a, Castellano2024a, Kumari2024a, Curti2024a}, and the green symbols are $z\sim1$\,--\,$3$ galaxies \citep{Stark2014a, Vanzella2016a, Vanzella2017a, Vanzella2020a, Iani2023a, Mainali2023a}. Gray symbols indicate local samples \citep{Senchyna2017a, Senchyna2019a, Berg2019a, Wofford2021a, Schaerer2022a}. Our CSFGs stand out as amongst the highest C$_{43}$-ratio sources with C$_{43}>1.6$ (excluding J1032+4919 with C$_{43}=0.7\pm0.5$), along with many of reionization-era \ion{C}{4} emitters.
\textit{Right}: C$_{43}$ \textit{vs.} $O_{32}$. The dashed orange line represents a linear model fitting to all samples, indicating a positive correlation between C$_{43}$ and O$_{32}$ ratios. We highlight that the majority of high O$_{32}$ ($>$10) sources exhibit C$_{43}$ ratios greater than 0.75 as the thresholds set the majority of strong LyC leakers with $f_{\rm esc,LyC}>0.1$.}
\label{fig:C43}
\end{figure*}

\subsection{Lyman-Continuum Escape}
High-redshift galaxies are thought to be significant contributors to reionization. Recent JWST observations during the EoR have, in particular, unveiled a higher abundance of galaxies than expected and suggest an elevated ionizing photon production rate \citep[e.g.,][]{Finkelstein2023a, Finkelstein2024a, Fujimoto2023a, Atek2024a}. However, the unknown LyC escape fraction from these galaxies — inherent limitation on the direct detection of LyC radiation during the EoR — complicates assessing the ability of these sources to supply ionizing photons to the IGM \citep{Munoz2024a}. Instead, local analogs of these galaxies serve as the best laboratories for understanding LyC escape of reionization-era galaxies.  

Our CSFGs are too close to observe leaking LyC photons directly with existing observing facilities. Instead, Ly$\alpha$ emission of our CSFGs, an indirect indicator of LyC escape, were targeted in previous HST observations \citep{Jaskot2017a, Izotov2020a}. In four of our sources (J0159+0751, J1205+4551, J1355+4651, and J1608+3528), Ly$\alpha$ emission lines are very strong (EW $>$160\AA) with narrow peak separations between the Ly$\alpha$ double peaks ($\lesssim300$ km\,s$^{-1}$). In contrast, J1032+4919 presents an exceptionally low Ly$\alpha$ emission of EW $=5.2$\AA\ with a triple-peaked line profile at the bottom of much broader Ly$\alpha$ absorption. \cite{Izotov2020a} provide the LyC escape values for three of our targets (J0159+0751, J1205+4551, and J1355+4651) based on the relation between Ly$\alpha$ peak separations ($V_{\rm sep}$) and LyC escape ($f_{\rm esc,LyC}$): Eq.(2) in \cite{Izotov2018b}. For J1608+3528, we separately estimated $f_{\rm esc,LyC}$ from its $V_{\rm sep}$ value given in \cite{Jaskot2017a} using the \cite{Izotov2018b} calibration. Based on the Ly$\alpha$-inference on LyC escape, these CSFG targets are promising LyC-leaker candidates with their derived $f_{\rm esc,LyC}$ values from 9 to 31\%. We note that the $f_{\rm esc,LyC}$ measurement is not given for J1032+4919 due to its complex Ly$\alpha$ line profile and the presence of a broad Ly$\alpha$ absorption. The Ly$\alpha$-related properties and derived LyC escape fraction measurements are listed in Table \ref{tab:lya}. In the following discussion on the $f_{\rm esc,LyC}$ measurements of our sources, we refer to the Ly$\alpha$-inferred-$f_{\rm esc,LyC}$ values.

With the presence of intense \ion{C}{4} emission in our sources, we examine their \ion{C}{4}/\ion{C}{3}] (C$_{43}$) ratios, a proposed LyC indicator \citep{Schaerer2022a}. In the left panel of Figure \ref{fig:C43}, we explore the relationship between the C$_{43}$ ratio and the LyC escape fraction of our sample with the known local LyC leakers \citep{Schaerer2022a}. Four galaxies among our targets show high \ion{C}{4}/\ion{C}{3}] (C$_{43}\geq1.6$) ratios with the exception of J1032+4919 (C$_{43}=0.7$), suggesting them as promising strong LyC-leaker candidates ($f_{\rm esc,LyC}>0.1$) based on the diagnostic presented in \cite{Schaerer2022a}. We omit J1032+4919 due to the complex Ly$\alpha$ line profile with weak triple peaks under the strong damped Ly$\alpha$ absorption. It is worth noting that triple-peak Ly$\alpha$ line profiles are often observed in LyC leakers \citep{Rivera-Thorsen2017a, Izotov2018b, Vanzella2018a}. However, the emergent Ly$\alpha$ flux and the Ly$\alpha$ escape fraction ($f_{{\rm esc,Ly}\alpha}$) of J1032+4919 are the lowest of the sample with the presence of a broad Ly$\alpha$ absorption. Observationally, $f_{{\rm esc,Ly}\alpha}$ tends to be greater than $f_{\rm esc,LyC}$, which suggests that J1032+4919 is not likely to have high $f_{\rm esc,LyC}$ \citep{Flury2022b}. This is consistent with the fact that J1032+4919 presents C$_{43} = 0.7\pm0.5$, lower than the criteria defining strong LyC leakers.

The middle panel of Figure \ref{fig:C43} shows the relationship between the C$_{43}$ and metallicity, indicated by 12+log(O/H), compiling available literature values. Similar to EW(\ion{C}{4}), we cannot draw a clear correlation between C$_{43}$ and gas-phase metallicity due to the narrow distribution of metallicities of \ion{C}{4} emitters. However, our CSFGs stand out as amongst the highest C$_{43}$-ratio sources with C$_{43}>1.6$ (excluding J1032+4919), along with many of reionization-era \ion{C}{4} emitters. It is suggestive that these high C$_{43}$ sources are likely strong LyC-leaker candidates (LyC escape fraction $\geq 0.1$) following the C$_{43}$ criteria ($>0.75$) given in \cite{Schaerer2022a}.

O$_{32}$ line ratios have been linked to $f_{\rm esc,LyC}$ as an indirect indicator \citep[e.g.,][]{Chisholm2018a, Izotov2018b}. Particularly, in the Low-redshift Lyman Continuum Survey sample, a majority of high O$_{32}$ ratios ($>$10) galaxies are found as significant LyC leakers with $f_{\rm esc,LyC}>0.1$ \citep{Flury2022b}. \cite{Mingozzi2022a} compare ionization parameters estimated from different zones with a C$_{43}$ ratio and find that C$_{43}$ ratios are correlated with O$_{32}$-based ionization parameters from the \ion{C}{4} emitters in the CLASSY sample \citep{Berg2022a}. We show C$_{43}$ against O$_{32}$ in the right panel, including our targets with literature measurements. The dashed orange line represents a linear model fitting to all samples, indicating a positive correlation between C$_{43}$ and O$_{32}$ ratios, which is consistent with the findings from previous work. The estimated Pearson correlation coefficient is $r = 0.63$ with $p\sim3\times10^{-4}$. This correlation suggests that despite a significant scatter, galaxies with high C$_{43}$ ratios also tend to have high O$_{32}$ ratios.  Such a scatter may be due to the fact that C$_{43}$ and O$_{32}$ ratios reflect somewhat different ionization zones: high/very-high \textit{vs.} high ionization zones \citep{Berg2021a} in addition to the complexity of \ion{C}{4} emission properties (e.g., possible stellar contribution and/or resonant scattering).  Nonetheless, we highlight that it is evident that the majority of high O$_{32}$ ($>$10) sources exhibit C$_{43}$ ratios greater than 0.75 as the thresholds set the majority of strong LyC leakers with $f_{\rm esc,LyC}>0.1$ \citep{Flury2022b, Schaerer2022a}. Our CSFGs fall into this category with most of the high-redshift sample, suggesting that they are strong candidates for significant LyC leakers.

\begin{figure*}[th]
\centering
\includegraphics[width=0.7\textwidth]{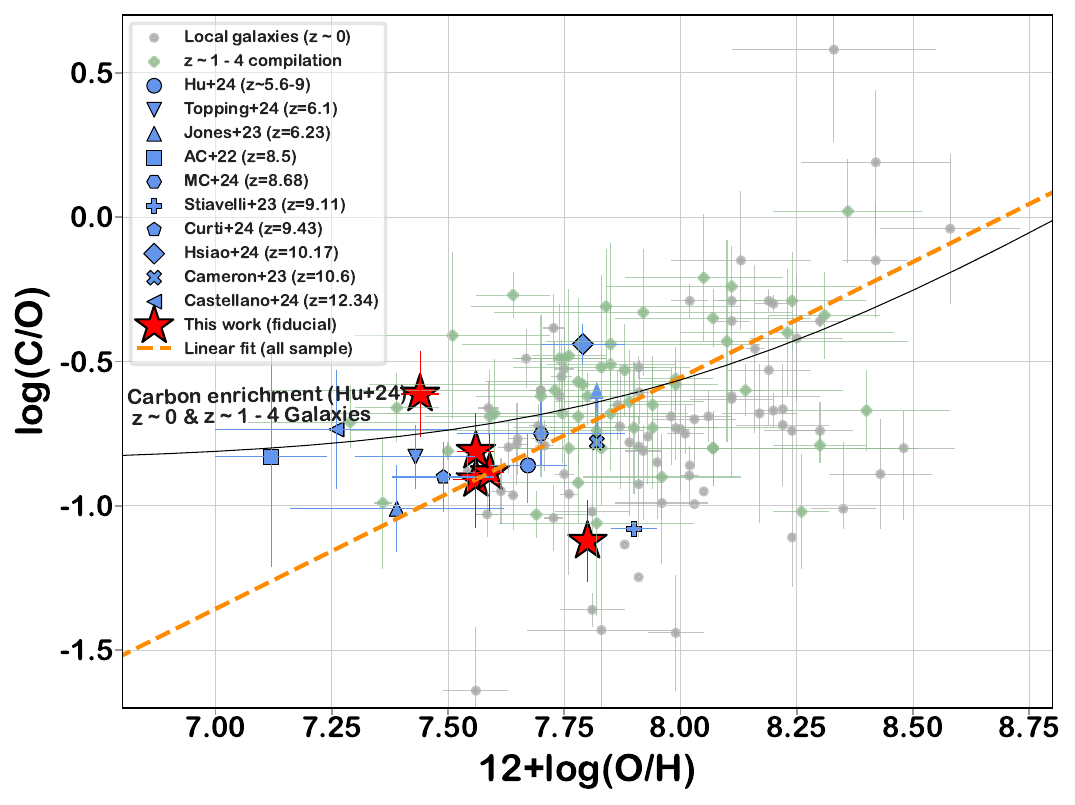}
\caption{C/O \textit{vs.} O/H. We present C/O abundances of our sources with literature values of local to reionization-era galaxies. The C/O abundances of local galaxies are marked as gray dots \citep{Senchyna2017a, Senchyna2021a, Pena-Guerrero2017a, Berg2019a, Ravindranath2020a, Rogers2023a, Izotov2023a}. The $z\sim1$\,--\,$4$ compilation is shown as green diamonds \citep{Erb2010a, Christensen2012a, Stark2014a, Bayliss2014a, James2014a, deBarros2016a, Steidel2016a, Berg2018a, Mainali2020a, Matthee2021a, Citro2024a, Iani2023a, Llerena2023a}. Reionization-era galaxies are presented as blue-color symbols \citep{Arellano-Cordova2022a, Jones2023a, Stiavelli2023a, Marques-Chaves2024a, Hu2024a, Topping2024a, Curti2024a, Hsiao2024b}. The black solid line shows the Carbon enrichment trend that is fitted for $z\sim0$\,--\,$4$ galaxies in \cite{Hu2024a} with the two-phase enrichment assumption \citep{Nicholls2017a}. Our targets feature significantly low C/O abundances, adding more scatter to the flat C/O trend at lower metallicity.  We show the best-fit linear model as the orange dashed line, accounting for all galaxy samples compiled in this work.} 
\label{fig:CO_abundance}
\end{figure*}

\subsection{Chemical Abundances}
Chemical abundances within these galaxies, particularly the C/O ratio, provide important clues about their nucleosynthetic origins and evolutionary processes \citep[see the review of ][]{Maiolino2019a}.  The C/O ratio is influenced by the production of carbon and oxygen in stars of different masses and metallicities.  Observations of C/O ratios in local SFGs help to understand the chemical evolution of galaxies and the delayed release of carbon from intermediate-mass stars \citep[e.g.,][]{Garnett1995a, Izotov1999a, Berg2016a, Berg2019a}.  Primary CNO nucleosynthetic processes dominate in metal-poor environments, leading to relatively constant C/O ratios. This flat C/O trend in metal-poor systems has often been reported in observations \citep[e.g.,][]{Berg2016a, Berg2019a, Izotov2023a}.  However, at higher metallicities, secondary processes become significant (either a secondary production or the delayed release of carbon), resulting in increased C/O ratios \citep[e.g.,][]{Garnett1999a, Henry2000a, Carigi2000a, Chiappini2003a, Nicholls2017a}.

In Figure \ref{fig:CO_abundance}, we present the C/O abundances \textit{vs.} the gas-phase metallicities of our targets compared to literature values over a wide redshift range. Local, $z\sim1$\,--\,$4$, and reionization-era galaxies are shown as gray, green, and blue symbols. The thin solid curve is the two-phase carbon enrichment trend, fitted for $z\lesssim4$ galaxies in \cite{Hu2024a}. Interestingly, the recent C/O measurements of reionization-era galaxies ($z\gtrsim6$) from JWST observations suggest lower C/O abundances of these early galaxies, compared to lower-redshift populations at $z\lesssim4$. \cite{Hu2024a} discuss that galaxies from the reionization era, on average, have lower carbon abundance in comparison to lower redshift galaxies \citep{Arellano-Cordova2022a, Jones2023a}. This trend is more clearly visible with additional C/O measurements of reionization-era galaxies \citep{Jones2023a, Stiavelli2023a, Marques-Chaves2024a, Topping2024a, Curti2024a}. Our sample galaxies feature their low C/O abundances comparable to or even lower than those of the reionization-era galaxies, adding more scatter to the flat C/O trend at lower metallicity.  This is somewhat contradictory to previous studies that report a flat trend of C/O with O/H at low metallicity \citep[e.g.,][]{Berg2016a, Berg2019a, Izotov2023a}. Also, a couple of reionization-era galaxies present C/O abundances higher than the majority of these galaxies, demonstrating a large scatter in C/O abundance against O/H down to a low metallicity range \citep{DEugenio2023a, Hsiao2024b}.

Overall, reionization-era galaxies and our sample fill the low-metallicity and low C/O abundance region, making it less favored to reproduce the empirical trend with the two-phase enrichment model. Instead, we fit a linear model of log(C/O) $= a + b\times$log(O/H) to all galaxy samples compiled in this work, including our sample in addition to the $z>6$ measurements from JWST observations. The best-fit linear model is presented as the orange dashed line with the best-fit parameters of $a=2.65\pm0.35$ and $b=0.80\pm0.09$, which accommodates well the low C/O samples. However, considering low-metallicity targets only (12+log(O/H) $<8$), it is difficult to tell if this prefers a linear trend rather than a flat one. We also note that the estimated C/O abundances can vary depending on the methods (refer to Section 3.2). Thus, this needs to be further tested with additional C/O measurements of low-metallicity galaxies, critically at 12+log(O/H) $\lesssim7.5$.

\section{Summary and Conclusions}
We analyze the HST COS and STIS spectroscopic observations of five local compact star-forming galaxies with extremely high [\ion{O}{3}]/[\ion{O}{2}] (O$_{32}>20$) ratios. Our targets present a wealth of rest-UV emission lines, including \ion{C}{4}, \ion{He}{2}, \ion{O}{3}], and \ion{C}{3}]. We measure emission-line properties and derive diagnostic line ratios to investigate ISM ionization conditions, ionizing sources, Lyman-continuum (LyC) escape, and chemical abundances. We summarize our major findings as follows.
\begin{enumerate}
    \item{Our targets feature intense \ion{C}{4} emission lines with EW $>$ 11\AA\ which require high-ionization condition in the ISM (log$U>-2$). We do not find obvious stellar contributions on the high-ionization lines of \ion{C}{4} and \ion{He}{2}. Also, we do not detect AGN features (e.g., \ion{N}{5}) in the rest-UV spectra.}
    \item{We use rest-UV emission-line diagnostics to examine ionizing sources of our targets. The emission-line diagnostics disfavor an AGN as the source of ionizing radiation. Interestingly, the diagnostics containing shock models (C3He2-O3He2 and C3He2-C4He2) suggest that shocks could be consistent with the emission-line ratios of our sources, which may also be true for the reionization era \ion{C}{4} emitters.  This suggests that the extreme ISM conditions of our sample require additional ionizing sources beyond the radiation emitted from young massive stars within standard stellar popualtions.}
    \item{Using data from the literature, we examine empirical correlations of \ion{C}{4} EWs and \ion{C}{3}] EWs with O$_{32}$ ratios. We find clear positive relations with O$_{32}$ although scatters become larger at higher O$_{32}$ ratios. In particular, our sample galaxies are positioned at the high ends of these relations along with reionization-era galaxies.}
    \item{Our CSFGs were targeted for their Ly$\alpha$ emission from previous HST observations \citep{Jaskot2017a, Izotov2020a}. Four of our sources (J0159+0751, J1205+4551, J1355+4651, and J1608+3528) present Ly$\alpha$ emission lines, that are very strong (EW $>$160\AA) with narrow peak separations between the Ly$\alpha$ double peaks ($V_{\rm sep}\lesssim300$ km\,s$^{-1}$) as well as significant fluxes at the Ly$\alpha$ centers. Such Ly$\alpha$ properties suggest that these targets are promising LyC-leaker candidates with their $V_{\rm sep}$-based $f_{\rm esc,LyC}$ values from 9 to 31\%. Exceptionally, J1032+4919 presents a weak Ly$\alpha$ emission (EW $=5.2$\AA) with a triple-peaked line profile at the bottom of much broader Ly$\alpha$ absorption, suggesting that this source is less likely to be a strong LyC leaker.}
    \item{Four galaxies among our targets show high \ion{C}{4}/\ion{C}{3}] (C$_{43}\geq1.6$) ratios with the exception of J1032+4919 (C$_{43}=0.7$), suggesting them as promising strong LyC-leaker candidates ($f_{\rm esc,LyC}>0.1$) based on the diagnostic presented in \cite{Schaerer2022a}. This is consistent with the inference from their Ly$\alpha$ properties. Compiling literature values with our sample, we find a positive correlation of C$_{43}$ ratios with O$_{32}$ ratios. Notably, the majority of high O$_{32}$ ($>$10) galaxies show C$_{43}>0.75$, suggesting them as potential strong LyC leakers ($f_{\rm esc,LyC}>0.1$).}
    \item{We derive relative C/O abundances from our sources, showing low log(C/O) values from $-1.12$ to $-0.61$, comparable to reionization-era galaxies. In the C/O -- O/H relation, reionization-era galaxies, and our CSFGs fill the low-metallicity and low C/O abundance region, which cannot be reproduced by the empirical trend with the two-phase enrichment model. Instead, these low C/O sources are well-accommodated with a linear model of log(C/O) $= 2.65 (\pm0.35) + 0.80 (\pm0.09)\times$log(O/H).}
\end{enumerate}

Our CSFGs with [\ion{O}{3}]/[\ion{O}{2}] ratios (O$_{32}>20$) presented in this study are currently the best analogs of reionization-era galaxies. We compile literature values, including the ones obtained from the latest JWST observations of reionization-era galaxies. We discuss galaxies with strong \ion{C}{4} emission in their UV spectra. In particular, the intense \ion{C}{4} with high \ion{C}{4}/\ion{C}{3}] ratios (C$_{43}$) of these galaxies suggest them as potentially significant contributors to the ionizing photon budget to reionization with $f_{\rm esc,LyC}>0.1$. As we find a significant correlation between C$_{43}$ and O$_{32}$ that most galaxies with high O$_{32}$ ratios ($>$10) show C$_{43}$ ratios above 0.75, those C$_{43}$ and O$_{32}$ thresholds are likely associated with significant LyC escape ($f_{\rm esc,LyC}>0.1$).  This positions the CSFGs, along with high C$_{43}$ and/or O$_{32}$ galaxies during the EoR, as strong candidates for LyC leakage.

Exploring public data from JWST PRISM observations in the JADES program \citep{Eisenstein2023a, Bunker2023b, DEugenio2024a}, we find that about 8\% of $z > 6$ galaxies with secure redshifts (15/189) present high O$_{32}$ ($>$ 10) ratios. This is a significant increase in the portion of such extreme galaxies, compared to the lower-redshift galaxies; e.g., spectroscopic surveys at $z\sim1-3$ rarely find O$_{32}>10$ galaxies \citep{Strom2018a, Papovich2022a}. Although these galaxies take less than 1/10 of entire reionization-era galaxy populations, they can be significant drivers in the process of reionization with their superior $f_{\rm esc,LyC}$ values. Further detailed analyses of these galaxies on their escaping LyC radiation would help in understanding the EoR.  While direct observations of reionization-era galaxies with JWST deliver rich information on their formation history and physical properties, it is fundamentally limited in measuring their escaping ionization photons due to IGM attenuation. In parallel with JWST observations of reionization-era galaxies, local analogs, like our targets, will remain as best probes to understand the physical conditions that are conducive to LyC escape.

\begin{acknowledgments}
We thank D. Berg, J. Chishomn, and M. Mingozzi for the useful discussion.
This research is primarily based on observations taken by the HST Cycle 28 program (GO 16213) with the NASA/ESA Hubble Space Telescope obtained from the Space Telescope Science Institute, which is operated by the Association of Universities for Research in Astronomy, Inc., under NASA contract NAS 5–26555. Support for this work was provided by NASA through a grant program ID HST-GO-16213. Some of the data presented in this paper were obtained from the Multimission Archive at the Space Telescope Science Institute (MAST). The HASP data of our observations (GO 16213) is available at MAST: \dataset[doi:10.17909/8xx0-y056]{\doi{doi:10.17909/8xx0-y056}}. 

Funding for the Sloan Digital Sky Survey V has been provided by the Alfred P. Sloan Foundation, the Heising-Simons Foundation, the National Science Foundation, and the Participating Institutions. SDSS acknowledges support and resources from the Center for High-Performance Computing at the University of Utah. SDSS telescopes are located at Apache Point Observatory, funded by the Astrophysical Research Consortium and operated by New Mexico State University, and at Las Campanas Observatory, operated by the Carnegie Institution for Science. The SDSS web site is \url{www.sdss.org}.

SDSS is managed by the Astrophysical Research Consortium for the Participating Institutions of the SDSS Collaboration, including Caltech, The Carnegie Institution for Science, Chilean National Time Allocation Committee (CNTAC) ratified researchers, The Flatiron Institute, the Gotham Participation Group, Harvard University, Heidelberg University, The Johns Hopkins University, L'Ecole polytechnique f\'{e}d\'{e}rale de Lausanne (EPFL), Leibniz-Institut f\"{u}r Astrophysik Potsdam (AIP), Max-Planck-Institut f\"{u}r Astronomie (MPIA Heidelberg), Max-Planck-Institut f\"{u}r Extraterrestrische Physik (MPE), Nanjing University, National Astronomical Observatories of China (NAOC), New Mexico State University, The Ohio State University, Pennsylvania State University, Smithsonian Astrophysical Observatory, Space Telescope Science Institute (STScI), the Stellar Astrophysics Participation Group, Universidad Nacional Aut\'{o}noma de M\'{e}xico, University of Arizona, University of Colorado Boulder, University of Illinois at Urbana-Champaign, University of Toronto, University of Utah, University of Virginia, Yale University, and Yunnan University.
\end{acknowledgments}

\vspace{5mm}
\facilities{HST(COS and STIS), Sloan (optical)}

\software{astropy \citep{Astropy2013a, Astropy2018a, Astropy2022a},
          matplotlib \citep{Hunter2007a},
          NumPy \citep{Harris2020a}, 
          SciPy \citep{Virtanens2020a},
          Photutils \citep{Bradley2023a},
          PyNeb \citep{Luridiana2015a}
}



\bibliographystyle{aasjournal}
\bibliography{references_all}



\end{document}